\documentclass[ArXiv,AXISWP,accept,moreauthors,pdftex,10pt,letterpaper]{Definitions/axis} 

\usepackage{caption} 

\firstpage{1} 
\pubvolume{xx}
\issuenum{1}
\articlenumber{5}
\pubyear{2023}
\copyrightyear{2023}

\Title{The life cycle of stars and their planets from the high energy perspective}

\Author{
Lia Corrales$^{1\dagger}$, 
Keivan G.\ Stassun$^{2\dagger}$, 
Tim Cunningham$^{3,4}$
Girish M. Duvvuri$^2$, 
Jeremy J. Drake$^3$,
Catherine Espaillat$^5$, 
Adina D. Feinstein$^6$, 
Elena Gallo$^1$,
Hans Moritz G\"{u}nther$^7$, 
George W. King$^1$, 
Marina Kounkel$^8$, 
Carey M. Lisse$^9$, 
Rodolfo Montez Jr.$^3$, 
David A. Principe$^7$,
Jes\'{u}s A. Toal\'{a}$^{10}$, 
Scott J. Wolk$^3$, 
Raven Cilley$^1$, 
Tansu Daylan$^{11}$, 
Margarita Karovska$^3$, 
Pragati Pradhan$^{12}$, 
Peter J. Wheatley$^{13}$
and 
Jun Yang$^{3,7}$
}

\AuthorNames{AXIS Stars \& Exoplanets Working Group}

\address{%
$^{1}$University of Michigan; 
$^{2}$Vanderbilt University; 
$^3$Harvard \& Smithsonian Center for Astrophysics; \\
$^4$NASA Hubble Fellow; 
$^5$Boston University; 
$^6$University of Colorado Boulder; \\
$^7$Massachusetts Institute of Technology; 
$^{8}$University of North Florida; \\
$^9$Johns Hopkins University, Applied Physics Laboratory;
$^{10}$National Autonomous University of Mexico (UNAM); \\
$^{11}$Washington University in St.~Louis;
$^{12}$Embry-Riddle Aeronautical University;
$^{13}$University of Warwick\\
$^\dagger$Corresponding authors: liac@umich.edu, keivan.stassun@vanderbilt.edu
}

\abstract{
One of the key research themes identified by the Astro2020 decadal survey is \textit{Worlds and Suns in Context}. The Advanced X-ray Imaging Satellite (AXIS) is a proposed NASA APEX mission that will become the prime high-energy instrument for studying star-planet connections from birth to death. This work explores the major advances in this broad domain of research that will be enabled by the AXIS mission, through X-ray observations of stars in clusters spanning a broad range of ages, flaring M-dwarf stars known to host exoplanets, and young stars exhibiting accretion interactions with their protoplanetary disks. In addition, we explore the ability of AXIS to use planetary nebulae, white dwarfs, and the Solar System to constrain important physical processes from the microscopic (e.g., charge exchange) to the macroscopic (e.g., stellar wind interactions with the surrounding interstellar medium). 
\emph{This White Paper is part of a series commissioned for the AXIS Probe Concept Mission; additional AXIS White Papers can be found at the  \href{http://axis.astro.umd.edu/}{AXIS website} with a mission overview \href{https://arxiv.org/abs/2311.00780}{here}}.}

\begin{document}
\tableofcontents
\listoffigures
\clearpage


AXIS is poised to become the prime instrument for studying the high-energy environments of stars and the corresponding effects on their respective planetary systems. 
Table~\ref{tab:AXIS} summarizes key AXIS qualities, relevant for studying stars and exoplanets, in comparison to modern X-ray observatories.
In Section~\ref{sec:birth}, we describe the state of the field when it comes to high energy characterization of protostars, from accretion processes to the effects of X-rays on planet formation. Section~\ref{sec:activity} describes the myriad of ways in which stellar activity, which produces both high-energy irradiation and coronal mass ejections, can influence the atmospheric properties, demographics, and habitability of planetary systems, including our own. Finally, Section~\ref{sec:death} describes how X-ray observations can help us understand the ultimate fate of Earth, through the study of planetary nebulae and white dwarfs. AXIS will provide leaps in understanding in all of these key science areas, tackling the Astro2020 theme: \textit{Worlds and Suns in Context} \citep{Astro2020}.

\begin{figure}[H]
\centering
\includegraphics[width=16 cm]{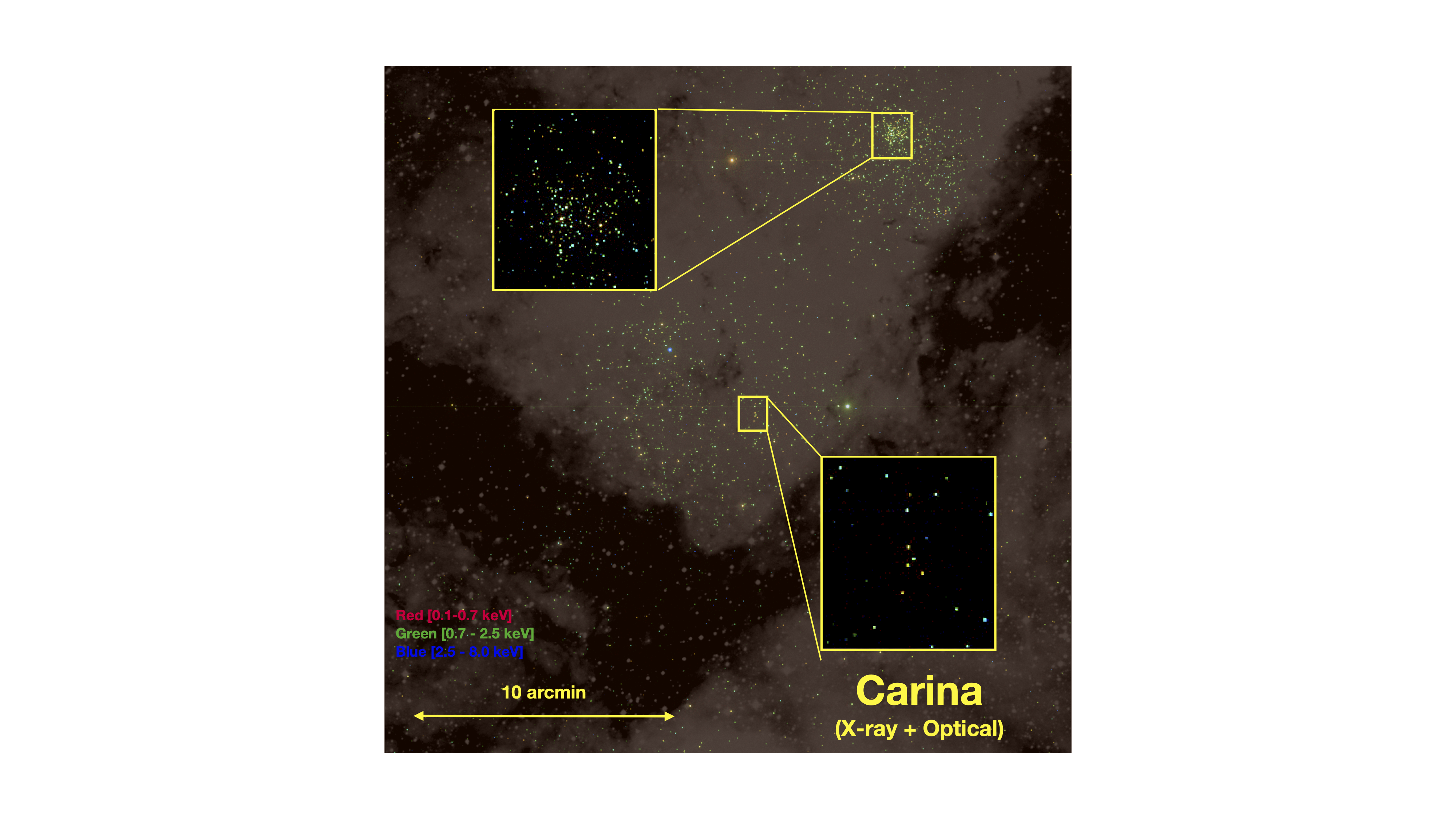}
\caption[Simulated AXIS observation of the Carina star-forming complex]{Simulated 50 ks three-color AXIS observation of point sources in a portion of the Carina star-forming complex. The X-ray image is overlaid with a DSS optical image (white), demonstrating the utility of X-rays piercing the molecular cloud material out of which stars are born. With stars having ages $1-6$~Myr, the entire Carina complex has one of the highest concentrations of massive stars in the Galaxy, in addition to more than 100,000 young, low-mass stars. A Chandra legacy survey of the Carina complex utilized 1.2 Ms over 22 pointings, yielding ~14,000 X-ray detections, 90\% of which had fewer than 50 counts \citep{2011ApJS..194....1T}. AXIS can perform even larger surveys in a fraction of the time thanks to its sharp PSF stability and
large collecting area across a wide field of view: A single 150 ks AXIS pointing in this portion of the star-forming complex can detect the $\sim 6000$ stars previously detected by Chandra and thousands of new, young, low-mass stars that were previously undetected, easily reaching M3-type dwarfs around the peak of the stellar IMF.}
\label{fig:Carina}
\end{figure}   

\begin{table}
\caption{Key AXIS observatory parameters}
\label{tab:AXIS}
\centering
\begin{tabular}{l|c|c|c}
    \toprule
    \textbf{Property} & \textbf{AXIS} &  \textbf{Chandra ACIS} & \textbf{XMM EPIC-pn} \\
    \midrule
    \textbf{0.5~keV Eff Area} (cm$^2$) & 3300 & 11 & 900 \\
    \textbf{PSF$^\dagger$ on-axis} (arcsec) & 1.5 & 0.8 & 15 \\ 
    \textbf{PSF$^\dagger$ at $9'$}  (arcsec) & 1.6 & 7.0 & 18 \\ 
    \textbf{Field of View} (arcmin$^2$) & 452 & 256 & 713 \\
    \textbf{$\Delta E$ (eV) Resolution @ 1 keV} & 60 & 100-150 & 80 \\ 
    \bottomrule
    \multicolumn{4}{l}{$^\dagger$As defined by the radius at which 50\% of the soft X-ray PSF is captured.} \\
\end{tabular}
\end{table}

\section{The birth of stars and planets}
\label{sec:birth}

The earliest stages of a star's life, from the initial collapse of a molecular cloud to the eventual fusion of hydrogen in a stellar core as it joins the main sequence, are described by a stellar classification system based on the shape of a star's infrared spectral energy distribution continuum---be it the cold broad
spectrum of a collapsing cloud, the mixed spectrum of a protostar + thick heated envelope, or the clearly
separated protostar + disk \citep{1987IAUS..115....1L}. The class system (Class 0, I, II and III) broadly describes major stages in a young star’s life which begins at Class 0 when a cold core begins to form as a star's natal molecular cloud clump begins to gravitationally collapse. Molecular clouds have some initial net rotation, and as they gravitationally contract, conservation of angular momentum results in a protostellar disk composed of gas and dust with a forming protostar at its center. This Class I stage is characterized by a protostar, shining via accretion of material from its surrounding environment (which includes a circumstellar disk and an envelope or shell formed from its parent molecular cloud). The Class II stage begins as a young star clears its surrounding environment and is characterized by an un-embedded, but still accreting, star-disk system. During the Class I and II stages, light from the central protostar shines on the disk, which can influence its chemistry and the eventual formation of planets. Eventually, a young star’s disk is dispersed (removed) leaving any already formed planets and the central pre-main-sequence star. It remains in this Class III stage until it joins the main sequence.

This sequence of stellar assembly involves several energetic processes – accretion, jets/outflows, and magnetic activity – that can produce X-ray emission, and therefore X-rays are key informants into the detailed physics of these processes. In turn, the X-rays produced by these stellar processes become an important source of high-energy insolation of the young planets that form in the circumstellar environment of the young star.

\subsection{Coronal activity, jets, and accretion of young stars}

Stars form through gravitational fragmentation and collapse of large molecular clouds. The initial infall is radial, but due to the conservation of angular momentum, the innermost region spins up as it contracts, and an accretion disk forms around it. In the earliest stages of star and planet formation, the central protostar is hidden within the dense cloud and can only be observed in the IR and longer wavelengths. For more massive stars, the evolution is so fast that most of the early phases remain invisible. In this summary, we concentrate on slower-forming stars with masses $< 3~M_{\odot}$. 

As accretion proceeds, the envelope, an outer shell of material from the parent molecular cloud of a star, becomes thinner. Many young stars have jets that eject mass and angular momentum and disperse parts of the envelope, making the star itself visible. This happens before stars reach the main sequence; the protostars are still contracting, but they already have convective envelopes and stellar activity. In this stage, protoplanets begin to form in the accretion disk, while material continues to fall onto the star, following magnetic field lines that connect the star to the disk. There are three distinct sources of X-rays during this period that we discuss in the following sections (see also the review by \citet{2022hxga.book...57S}).

\subsubsection{Magnetic Activity}

Magnetic fields drive stellar activity through tangling, snapping, and reconnection events in the protostar's atmosphere that heat X-ray-emitting plasmas and accelerate electrons to produce synchrotron radiation. While seemingly rare, even the youngest (Class 0) stage of a star's life demonstrates X-ray emission characteristic of magnetic activity \citep{2020A&A...638L...4G}. For other young stars, primarily those at Class I and later, occasional flares with large loop lengths are observed \citep{2005ApJS..160..469F}; those flares might occur along the magnetic field lines that connect the star and the disk. Stars that are still accreting (Class II or CTTS) exhibit X-ray luminosities that imply less magnetic field activity than later stages of young stars where the accretion disk has dispersed (Class III or WTTS), \citep{2005ApJS..160..401P}. The physical mechanism for this difference, thoroughly demonstrated in the Chandra Orion Ultradeep project \citep[COUP][]{2005ApJS..160..319G}, is still under debate two decades later. Stellar accretion might suppress the coronal X-ray emission by filling the corona with colder plasma or disturbing the structure of the magnetic field lines. However, it is still possible that this is an observational bias because Class II sources are more obscured and thus the soft X-rays they emit might be less observable. Given this uncertainty, future large surveys with AXIS that sample time-domain activity levels for a range of conditions in star-forming clouds (Figure~\ref{fig:Carina}) will provide a clear path for confirming whether or not magnetic activity is suppressed at earlier stages of star formation. AXIS’s small PSF, which is stable over a larger field of view than Chandra, enables surveys not just of the cores of star-forming regions, but also the outer, less dense, regions to cover a wide variety of conditions.

\subsubsection{Accretion shocks}

High-resolution observations of the He-like triplets of O VII and Ne IX show that this emission is produced in relatively high-density regions ($n_e \sim 10^{12}$~cm$^{-3}$), which are much higher than in the corona of a main sequence star ($n_e \sim 10^{10}$~cm$^{-3}$). Instead, they originate in the accretion shock, where material from a protostar's circumstellar disk falls onto the body of the star \citep{2002ApJ...567..434K,2010ApJ...710.1835B}. As the material falls onto the star, it is accelerated by gravity and reaches free-fall velocities of $300-500$~km/s (depending on the mass and
radius of the star). An accretion shock is formed, heating matter to soft X-ray emitting temperatures ($\sim 1-3 \times 10^6$~K) when the material impacts the stellar surface. The triplet spectral features from He-like elements likely arise from the cooling zone of that shock and from material that has escaped magnetic confinement, such as winds or outflows with densities higher than typically encountered in stellar coronae. Understanding the contribution of He-like ions to the total X-ray flux of young stars is one of the major model ingredients that will go into interpreting the X-ray flux of young stars observed with AXIS.

\subsubsection{Shocks in outflows: jets}

Young stars often shed angular momentum through outflows: from slow-moving, wide-angle disk winds to fast and highly collimated jets launched close to the star. In some cases, those jets can be traced to several pc from the star. X-ray emission can be seen from the termination shock where the jet runs into the interstellar medium \citep[e.g.][]{2003ApJ...584..843B,2005ApJ...626..272P,2006A&A...448L..29G},  from inner working surfaces in the shock \citep{2010A&A...517A..68B}, and also from the inner region very close to the launching region \citep{2007A&A...468..515G,2011ApJ...737...54B}.

Inner jets, in particular, have been shown to be X-ray emitters in a number of young stars. X-rays from shocks in outflows that have been resolved by Chandra are faint and located close, sometimes within an
arcsec (10s of au) of the central star. For younger stars that are still deeply embedded in molecular clouds, a jet that pierces an opening into the infalling envelope may be observable even while the central protostar is not, due to obscuration by its natal cloud. In those cases \citep[e.g. HH~154][]{2006A&A...450L..17F}, all the observed X-rays come from jet-induced shocks;  the observed emission is spatially offset from the central source and is so soft that it cannot come from the embedded star.

Stellar jets provide feedback on the molecular cloud, pierce holes that make the inner star visible to us, and carry away angular momentum that allows accretion to proceed. Understanding jets is absolutely crucial to make progress on several science questions identified in the Astro2020 decadal survey for “Interstellar Medium and Star and Planet Formation”. In particular, F-Q2a asks: “What processes are responsible for the observed velocity fields in molecular clouds?”. To answer this question, we need to quantify the mass and momentum that stellar jets inject into the clouds. For question F-Q3b: “How do protostars accrete from envelopes and disks, and what does this imply for protoplanetary disk transport and structure?”, we need to quantify the angular momentum removed from the system by winds and outflows and determine at what radius that occurs. Since X-rays probe the fastest outflow component of the jet, which are launched deep inside the gravitational well of the star, possibly even from the inner disk interface or the stellar surface, they are crucial to getting a complete picture of the stellar jets and their impact on star formation and molecular clouds.

\subsection{Case study of stellar jets: DG Tau}

\begin{figure}[H]
\centering
\includegraphics[width=15 cm]{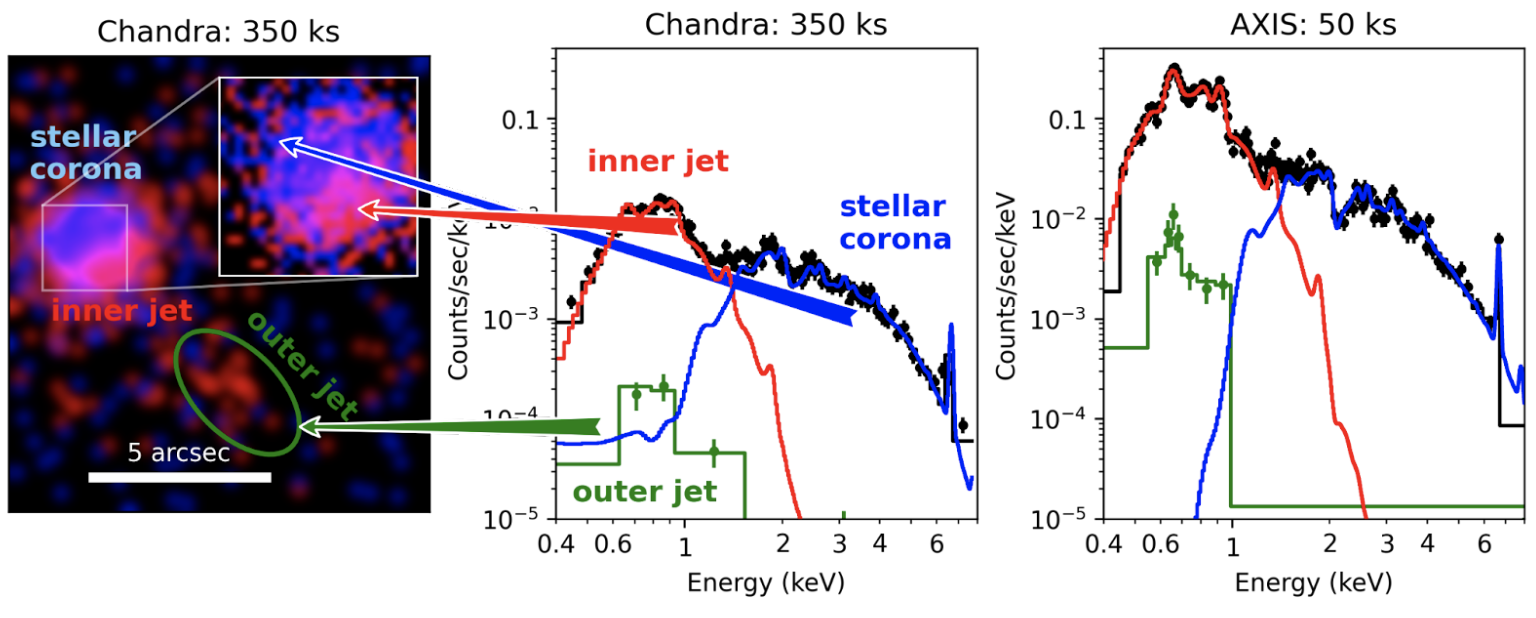}
\caption[The classical T-Tauri star DG~Tau in X-rays]{The CTTS DG~Tau in X-rays. The left panel shows a 350~ks Chandra image combining multiple observations taken in 2010. The image is color-coded with photons below 1.2~keV shown in red and photons above 1.2~keV in blue. Three spatially distinct components are visible. The DG Tau corona is absorbed, and only photons $> 1.2$~keV are visible (blue) at this location. There is a softer inner jet component (red), that is offset about 0.3~arcsec from the stellar position, but due to the resolution of this image, the two components overlap. Even with the 1~arcsec resolution of Chandra, the two components can be spectrally resolved by fitting with two different temperatures and absorption components (black data with red and blue model components in the middle panel). There is also an outer jet component that is several arcsec (650~AU) away from the central star. We expect substructure in this region, but given the low count rate in Chandra, we only know that it is spatially extended. This feature will be resolved with AXIS, and a spectrum can be extracted separately (green). The right panel shows a simulation for a 50~ks AXIS observation of DG Tau based on the models fitted to the Chandra data. In both the middle and the right panels, the jet (green) spectrum is binned to 10 counts/bin, and the stellar/inner jet spectrum to 30 counts/bin. The superior collecting area of
AXIS will allow us to fit abundances and multi-temperature models, search for substructure, and look for
proper motion of the inner and out jet components in a fraction of the observing time required by Chandra.
}
\label{fig:DGTau}
\end{figure}

Class II sources are typically evolved star-disk systems that have cleared out their surrounding molecular environment. Observing Class II sources with jets provides an opportunity to investigate jet-launching mechanisms by revealing structure closer to the launching region. The best-studied object of this type is the CTTS DG Tau, which has 450 ks of cumulative observations with Chandra \citep{2005ApJ...626L..53G, 2008A&A...478..797G,2011ASPC..448..617G}, and reveals two spatially distinct components that are X-ray emitters in the outflow. Proper motion has been observed from X-ray jets \citep[e.g.][]{2012A&A...542A.123S}, but for DG Tau this question is unsettled. Figure~\ref{fig:DGTau} shows a 350~ks Chandra image of DG~Tau, which was obtained in a single year. This image resolves this system into three components: the young stellar corona, an inner jet region, and an outer jet region.

\subsubsection{The inner jet of DG~Tau}

The inner jet component is barely 0.2'' offset from DG Tau itself (30-50~AU), but its spectrum is much softer than the star because the star is deeply embedded, but the jet itself is not. This allowed \citet{2008A&A...488L..13S} to compare the position of the soft and hard photons and measure the offset, despite the fact that it is much smaller than the size of the Chandra PSF. Similar techniques can be employed with AXIS, but thanks to the much larger collecting area, measurements like this will no longer be limited by the low count rate. Together with the implementation of sub-pixel resolution algorithms, such as those developed for Chandra-ACIS, AXIS will thus be able to probe the evolution of the jet over time through repeat observations of this target over the course of the mission lifetime. 

Different hypotheses have been presented to predict the time revolution of DG~Tau’s inner jet. \citet{2010A&A...511A..42B} present a model of a diamond shock, and \citet{2014ApJ...795...51G} suggests a model where hydrodynamic pressure of the outer wind component recollimates the inner, faster, less dense region and focuses a standing recollimation shock. In these scenarios, the shock position should be mostly steady over time, but the shock intensity should change with changes in the accretion (and thus mass outflow) rate. Only AXIS has the collecting area and angular resolution to perform these measurements repeatedly and for multiple stars.

While DG Tau is so far the only known example where the jet shock is spatially resolved in imaging, the same spectral shape consisting of a hard, highly absorbed stellar corona and a much softer, much less absorbed shock component is also seen in GV~Tau and DP~Tau \citep{2007A&A...468..515G}. It seems likely that these objects also have X-ray-emitting jets. Unfortunately, with the major loss of the Chandra ACIS soft energy effective area, those observations are no longer feasible with Chandra, even for DG~Tau, which is the brightest of the known X-ray sources that exhibit evidence for a corona-jet system. AXIS is the only Probe-class X-ray telescope under design in the next decade that would be able to prove or disprove this idea directly through deep imaging.

\subsubsection{The outer jet of DG~Tau}

In addition to the inner jet shock, DG Tau also shows an even fainter, outer region with X-ray emission. This region is clearly resolved by Chandra and will be clearly resolved by AXIS. Unfortunately, 350 ks of Chandra observation yield only $\sim 50$ photons from this region. Based on observations in the UV and optical, where the signal is better and HST can resolve finer details, we know that there is considerable substructure, including a bow shock with hot and cold materials. However, the X-rays must come from a different component in the shock, since neither the temperatures nor the velocities of the components seen in the IR, optical, and UV are sufficient to explain the fitted X-ray temperature of 0.2~keV. Assuming that a bow shock or termination shock is responsible, the mass loss rate seen in the X-rays is about $10^{-10} M_{\odot}/{\rm yr}$ \citep{2009A&A...493..579G}, a factor of 1000 below the mass flux seen in the optical \citep{1995ApJ...452..736H}. At the same time, the velocity must be at least 300 km/s to explain the observed X-ray spectrum. With AXIS, we will be able to obtain an exposure deep enough to pinpoint the location of the X-ray emission and associate it with the jet structure. We can also repeat the observations over the lifetime of the AXIS observatory to follow the proper motion.

There are two possible explanations for the X-ray emission in the outer jet. Either the gas has been heated as it passed through a standing shock close to DG~Tau and is now adiabatically and radiatively cooling, or it is going through a shock wave at its current location. In the latter case, that shock wave could be caused by an obstacle such as a previous, slower outflow blob, or it could be a shock wave traveling along the jet. AXIS observations of DG Tau can distinguish all of those possibilities. If the gas has previously been heated and is cooling adiabatically, we expect the jet to have a cone opening angle. We know that the outer shock is resolved with Chandra, but we have insufficient signal to constrain its shape. AXIS will deliver an order of magnitude more counts and an arcsec-scale PSF -- enough to detect an opening angle of 5 degrees at 5 arcsec from the central source. Without expansion, the density of the emitting material is so high that it cools much faster, and no X-ray emitting material could reach beyond $\sim 1$~arcsec (140 AU, well past the system's proto-Kuiper Belt. Thus, if AXIS finds the outer X-ray jet to be highly collimated, the jet material must be heated in place by a shock. AXIS imaging can pinpoint the shock location to see if it is associated with a known, cooler structure seen in other wavelengths, which the X-ray jet runs into, as suggested by infrared observations \citep{Maurri:2014}. AXIS spectroscopy (Figure~\ref{fig:DGTau}) determines the temperature of the plasma, and thus the shock speed (Chandra indicates about 300~km/s; however, that fit is very uncertain since it is based on only $\sim 40$ photons). AXIS observations spaced by a year or more would also reveal the proper motion. At the distance of DG~Tau, 300~km/s corresponds to 0.5~arcsec/year, and with just 50~ks of AXIS observing the centroid of the jet emission could be determined through sub-pixel resolution algorithms. A simple comparison between proper motion and shock speed will show if the shock plows through the moving jet material or if the material passes a stationary shock wave (e.g., caused by an obstacle or the magnetic field).  In other words, AXIS will finally reveal what physical
mechanism powers X-rays from young stellar jets.

\subsubsection{Prime targets for AXIS}

DG Tau is just one example of a young star with a resolved X-ray jet. We only discuss it in detail because it has the deepest Chandra exposure of any other system, and thus we can make the most detailed predictions of what parameters AXIS will be able to measure. Overall, the AXIS PSF and larger collection area will allow studies more detailed than what Chandra has done on that single jet for a larger sample of sources. Table~\ref{tab:youngstars} gives examples of young stars with known X-ray jets showing the variety of conditions and environments that can be probed with AXIS.

\begin{table}[H]
\caption{Prime targets for observing protostellar jets}
\label{tab:youngstars}
\centering
\begin{tabular}{c|c|c}
    \toprule
    \textbf{Object} &  \textbf{Type} & \textbf{Chandra ref} \\
    \midrule
    DG~Tau & CTTS with resolved inner and outer jet emission & see text \\
    HH 168 & Large scale diffuse emission with knots & \protect{\citep{2009A&A...508..717S}} \\
    Cepheus A & Inner jets, source itself embedded & \protect{\citep{2009&A...508..321S}} \\
    L 1551 IRS5 (= HH 154) & Inner, stationary jet component and outer knot with proper motion & \protect{\citep{2002A&A...386..204F,2003ApJ...584..843B,2011A&A...530A.123S}} \\
    HH 2 & & \protect{\citep{2001Natur.413..708P}} \\
    RY Tau & Similar to DG Tau & \protect{\citep{2018ApJ...855..143S}} \\
    HD 163296 & Herbig AeBe star, more massive than CTTS & \protect{\citep{2005ApJ...628..811S,2013A&A...552A.142G}} \\
    \bottomrule
\end{tabular}
\end{table}

Beyond known sources, AXIS will be a discovery engine for new stellar jets. Its PSF remains concentrated over a much larger field-of-view than Chandra's PSF in the 8.3'$\times$8.3' ACIS
FOV.  This enables searches for resolved extended emission over much larger regions, including those with dozens to hundreds of other young stars with potential jets, in a single observation. At the same time, the increased effective area collects enough photons to securely identify jets in star-forming regions and measure jet speeds via shock temperatures. Currently, the number of jets known in optical and IR vastly outnumber the known X-ray-detected jets. Given the low soft X-ray sensitivity of Chandra, we cannot test whether this means that special conditions are needed to form the fast, X-ray-emitting outflow components or whether our sensitivity is simply too low to detect them in most cases. 

With AXIS, we can answer the question: Does every stellar jet have an inner fast component? If so,
that opens the door to a unified theory of how angular momentum is removed in the innermost regions
of the disk, allowing for ongoing accretion and disk clearing.

\subsubsection{Absorption by accretion funnels or the inner disk}

In addition to probing hot, X-ray-emitting gas, AXIS can also use X-ray absorption as a tool to study the environments of planet formation. Soft X-rays are absorbed by intervening material, such as molecular clouds or circumstellar disks, changing the shape of the observed spectrum. Although this absorption is typically quantified as ``equivalent hydrogen column density,'' it is actually dominated by photons absorbed by the inner shell electrons of metals, such as C, O, or N \citep{2000ApJ...542..914W}.

\subsubsection{Broad-band absorption}

When absorption by individual elements cannot be resolved, it is useful to assume a set of standard abundances and simply calculate the total gas density that would provide the correct density of metals. X-rays are absorbed, regardless of whether the metal is in atomic form, bound in a molecule, or in a small grain. X-rays are fully absorbed (``gray absorption'') only if the material is bound in large grains, pebbles, or planets. In this case, the spectral shape stays the same, but the observed flux is lower. In contrast, in the optical and IR, only dust grains contribute to the reddening. Thus, comparing the X-ray absorbing column density and optical reddening gives us a handle on the gas-to-dust ratio, which we know must vary
from the $\sim$100:1 of the ISM and moleuclar clouds to the $\sim$1:1 found in our mature solar system. Thus this
measurement is directly relevant to Astro2020 decadal survey question F-Q4b: ``What is the range of physical environments available for planet formation?''

From the source at the star itself, line-of-sight X-rays pass through different environments: The corona, the accretion streams, the disk atmosphere, the surrounding envelope, the large-scale molecular cloud, and lastly, the ISM between the star-forming region and us. All of those regions cumulatively contribute to the X-ray absorbing column density, making the interpretation of any result ambiguous. One of the most promising avenues is to study objects with time-variable column densities. For example, in AA Tau, \citep{2007A&A...462L..41S,2007A&A...475..607G} the accretion funnel passes through the line-of-sight periodically every 8.5~days, increasing the N$_{\rm H}$ by about $8 \times 10^{21}$~cm$^{-2}$. Compared to optical/IR data, this number indicates an elevated gas-to-dust ratio, compatible with the idea that significant dust evolution happens in the inner disk. For AA Tau we also have a probe of the outer material. In 2013, the star began to dim by up to 2 mag in the V band for periods of months or weeks. H$_2$ profiles indicate that the source dimming is located at $> 1$~AU. At this location, the ratio between X-ray absorbing column density and reddening matches observations in the ISM \citep{2015A&A...584A..51S}, so the grains through which the line of sight passes have not evolved.

A similar sudden dimming has been observed in CTTS RW Aur A \citep{2015A&A...584L...9S}, a member of the RW~Aur binary system, that can be resolved with Chandra or AXIS. At its maximum, the measured N$_{\rm H}$ increased by more than two orders of magnitude and interestingly, \citet{2018AJ....156...56G} find indications for an extremely high Fe abundance in the emitting plasma at that time. That iron has to be recently accreted, possibly due
to the breakup of an inwardly migrating differentiated planetesimal [Lisse 2022 REF], or because the grains migrating inward in the disk were trapped in a pressure trap zone that was suddenly released as the structure of the inner disk changes \citep{2019ApJ...871...53G} with obvious implications for planet formation in the inner regions of this system. These two examples show the range of science that detailed X-ray monitoring of a wide range of young stellar objects can deliver. With its arcsec-scale PSF over a
wide field of view combined with a large collecting area, AXIS will be able to sample a much larger
number of young stars than Chandra or XMM could and will lead to more objects of the type of AA Tau or
RW Aur A being detected to study, as well as reveal new phenomena in young stars that we have not yet
thought about.

\subsubsection{Detecting absorption signatures from individual elements}

Young stars with planet-forming disks (e.g., Class I and Class II) show a characteristic increase in absorbing column (N$_{\rm H}$) as a function of increasing circumstellar disk inclination \citep{2005ApJS..160..511K}. For example, the viewing geometry of an edge-on disk (i.e., a star-disk system whose central star must shine X-rays through the midplane of the disk before arriving at our detectors) results in much larger absorption than those viewed face-on. This trend was identified using a small number of sources whose disk inclinations were strongly constrained by HST when light from background nebulosity revealed the disk silhouette.

\begin{figure}[H]
\centering
\includegraphics[width=12 cm]{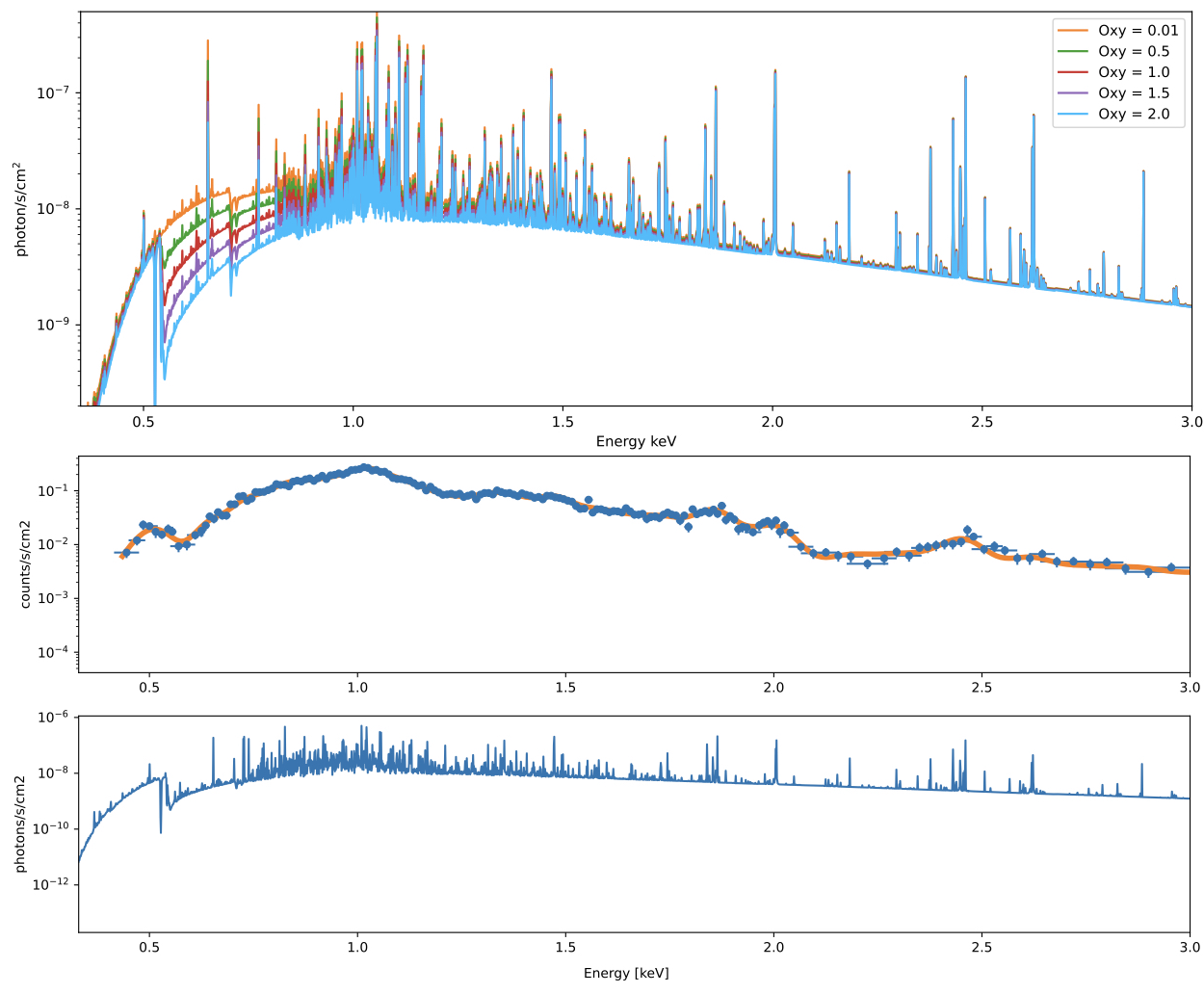}
\caption[Simulated absorption spectra from planet-forming disks]{\textit{(Top)} Five one-temperature absorbed plasma models differing only in oxygen abundance of the absorber (colored lines). The oxygen absorption feature at $\sim 0.55$~keV can be used to strongly constrain the oxygen content in nearby young planet-forming disks. \textit{(Middle and Bottom)}  A 150~ks AXIS simulation using a typical flux and absorbing column density for nearby young stars. The underlying model used to simulate the data (bottom) is an absorbed multi-temperature thermal plasma with an absorber oxygen abundance of 2~Solar. The large effective area of AXIS will allow a determination of oxygen abundance for nearby young planet-forming star-disk systems.}
\label{fig:diskabs}
\end{figure}

The Atacama Large Millimeter Array (ALMA) has revolutionized our understanding of circumstellar disks, in part due to its extreme sensitivity and impressive angular resolution capabilities. ALMA is sensitive to the long wavelength thermal dust emission of large mm to cm-sized grains in circumstellar disks, capable of imaging and spatially resolving hundreds of nearby young stars. It can also image the
disks in molecular hydrogen tracers such as isotopologues of CO and a plethora of molecular rotational
lines revealing that disks are chemically and morphologically complex \citep{2018ApJ...869L..41A, 2021ApJS..257....5Z}.  Depending on the viewing geometry of the disk to our line of sight, disk-absorption of X-rays from the central star can be used to constrain the abundance of oxygen (Figure~\ref{fig:diskabs}), the 3rd most abundant atom in the galaxy and a vital element
in water, rocks, and simple organics like CO, H$_2$CO, CH$_3$OH---and thus critical for the chemical evolution of disks,
planets, and life as we know it  \citep{2020ApJ...903..124B,2015ApJ...804...40G,2018MNRAS.473L..64E}.

In order to constrain the abundance of oxygen using X-ray spectroscopy, it is very important to collect enough signal at energies between 0.3~keV and 2~keV where the signature of oxygen absorption is highest. To fully sample the transition of X-ray absorption starting at 0.55 keV, it is important to collect a large number of X-ray events in the 0.30--0.65 keV region. Unfortunately, given the typical x-ray fluxes of
nearby young pre-MS stars, modern X-ray telescopes like XMM-Newton and Chandra lack the
sensitivity at 0.30--0.65 keV to confidently detect this specific oxygen absorption feature given the typical flux of nearby young pre-MS stars. An instrument like AXIS, with a high soft effective area and stable PSF over a large field of view, will allow us to identify this characteristic oxygen absorption for many sources when viewing crowded young stellar clusters.

\subsection{The effect of stellar activity on planet formation}

High-energy radiation from young stars is expected to play a critical role in energy deposition into the
surrounding protoplanetary disk from which planets form, as X-rays can penetrate through regions of a
UVIS opaque, X-ray optically thin PMS accretion disk and ionize and heat disk molecular gas. For example, the H$_{13}$CO$^{+}$  line strength has been observed to vary, and this behavior has been linked to changes in X-ray emission, presumably through an X-ray flare that enhanced the HCO$^{+}$ abundance in the disk \citep{Cleeves2017}.  More recently, SO, linked to X-ray emission in circumplanetary disks \citep{Law2023}, has been found to trace an
embedded protoplanet in a disk \citep{Young2021}, opening a new window of study of the chemical properties of the disk.  

Two major areas where AXIS can have an impact on our understanding of protoplanetary disks and planet formation are photoevaporation and disk ionization, which we discuss next.

\begin{figure}[H]
\centering
\includegraphics[width=12 cm]{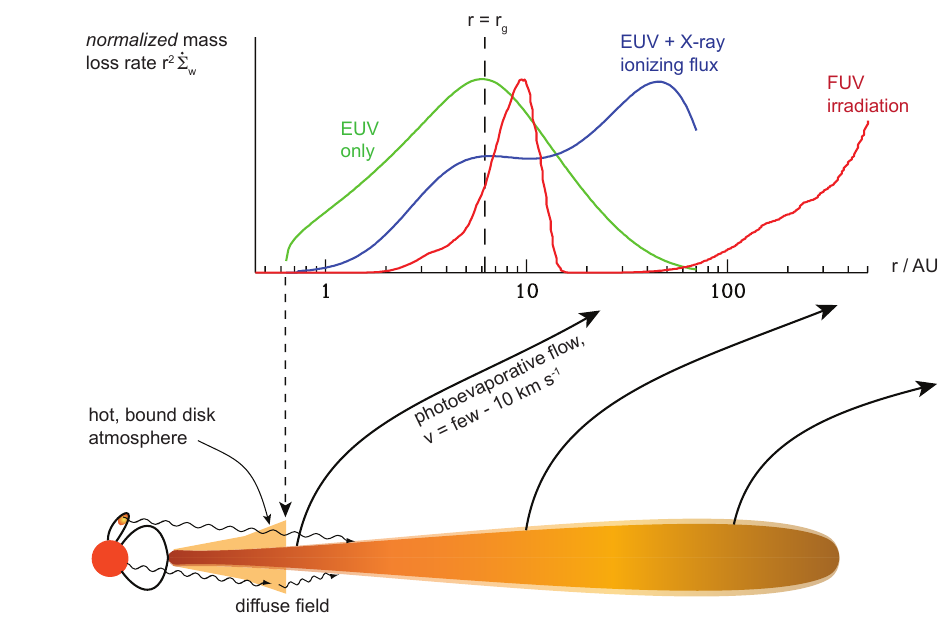}
\caption[Illustration of disk photoevaporation due to high-energy stellar radiation]{Illustration of disk photoevaporation due to high-energy stellar radiation (from \citet{Armitage2011}).  Relative mass loss rates with radius are shown for photoevaporation driven by EUV \citep{Font2004}, FUV \citep{Gorti2009}, and X-ray plus EUV radiation \citep{Owen2010}.}
\label{fig:diskphotoevap}
\end{figure}

\subsubsection{Photoevaporation}

The origin of planets is intricately tied to the evolution of protoplanetary disks. Significant gas must be present in the inner disk for gas giant planets to form, so the lifetime of the disk places an upper limit on the timescale for giant planet formation. The longevity of gas in the disk is dictated by the rate at which gas is eroded by photoevaporative winds created by high-energy ionizing radiation from the central star. However, there are no robust observational constraints on the connection between high-energy radiation and the dissipation of gas in disks. This lack of observational constraints has led to uncertainty about the efficiency of photoevaporation during the critical 1–10 Myr period when planets are expected to form. 

High-energy X-ray, EUV, and FUV radiation can ionize the upper disk layers, causing the gas to become unbound and escape, producing a thermally driven photoevaporative wind (Figure~\ref{fig:diskphotoevap}). Models of EUV photoevaporation predict low mass loss rates of $\sim 10^{-10} M_{\odot}/{\rm yr}$ \citep{Font2004, Alexander2006} while models that include X-ray or FUV emission achieve high mass loss rates up to $\sim 10^{-8} M_{\odot}/{\rm yr}$ \citep{Gorti2009, Owen2010}. 

Initially, in the first $\sim 10^5$ yrs, viscous evolution and radially inward mass flow dominate disk evolution, leading to a decrease in the mass accretion rate with time as disk material is depleted onto the star (i.e., less gas leads to a slower accretion rate). Once the rate of mass accretion inward through the disk equals the outward mass loss rate above, though, the inner disk is no longer able to be replenished
and its inventory is rapidly depleted, leaving behind an inner disk hole \citep{Clarke2001}.

This rapid disk clearing onset is reached later for the EUV photoevaporation model, since it takes longer for the disk accretion rate to decline to the lower $\sim 10^{-10} M_{\odot}/{\rm yr}$ level \citep{Hartmann1998}. 
The average mass accretion rate of young stars with disks is $\sim 10^{-8} M_{\odot}/{\rm yr}$ \citep{Hartmann1998}, which is the point at which X-ray- or FUV-driven
photoevaporation models are expect to dominate disk clearing.  However, the existence of young stars surrounded by substantial disks with mass accretion rates $ < 10{-10} M_{\odot}/{\rm yr}$ indicates that X-ray or FUV-driven photoevaporation may not always dominate, suggesting that EUV-driven photoevaporation may play a role \citep{Ingleby2011}; why this is the case is not understood. The main obstacle in distinguishing between photoevaporation models is the lack of observational constraints on the high-energy emission of stars, which AXIS will provide.

\begin{figure}[H]
\centering
\includegraphics[width=12 cm]{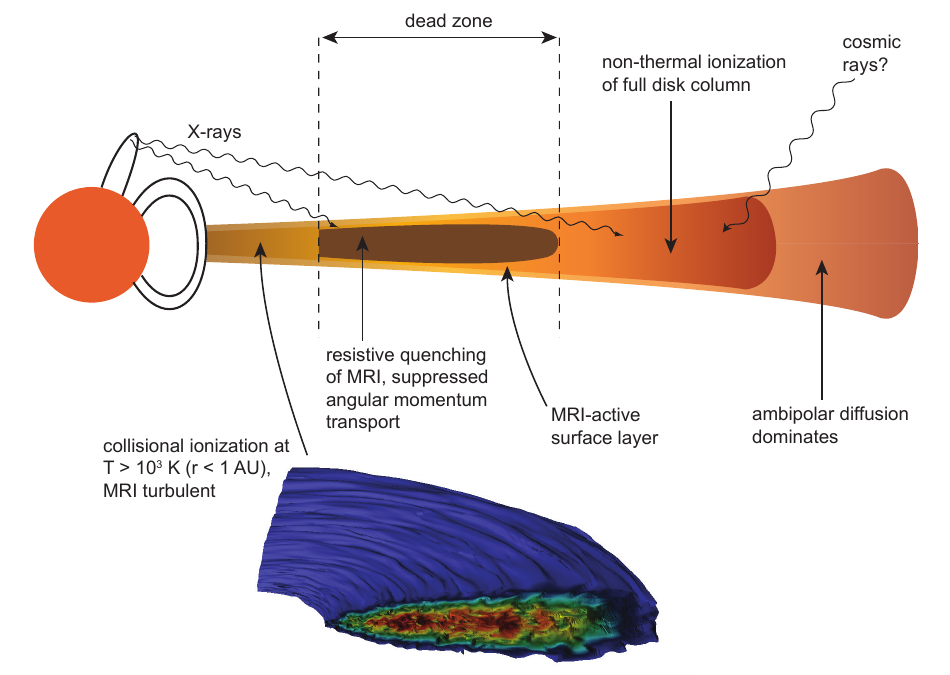}
\caption[Illustration of angular momentum transport in a protoplanetary disk]{Illustration of angular momentum transport due to the MRI in a protoplanetary disk (from \citet{Armitage2011}). There is a ``dead zone'' deep in the disk where X-rays cannot penetrate; this is covered by an ``active layer'' where X-rays ionize the disk and accretion occurs. The inner disk within 1~AU is fully MRI turbulent, as shown in the bottom simulation of density isocontours.}
\label{fig:MRI}
\end{figure}

\subsubsection{Disk Ionization}

Accretion in protoplanetary disks is essential for planet formation since the material in the disk must move inward, and some of this material goes into forming planets. However, we do not know the source of the angular momentum transport mechanism that leads to accretion. More than 30 years ago, magnetorotational instability (MRI) was proposed to explain how turbulent viscosity could cause accretion in weakly magnetized differentially rotating disks (Figure~\ref{fig:MRI}; \citet{Balbus1991}). However, the disk must be sufficiently ionized for the MRI to be active, and there are likely ``dead zones'' in regions of the disk that are not ionized \citep{Gammie1996}. Cosmic rays cause too little ionization to produce the MRI \citep[e.g.,][]{Cleeves2013}, but X-rays can ionize disks enough for the MRI to take hold in at least the uppermost disk layers \citep{Glassgold1997}.

However, corresponding observational support for the MRI is still elusive. ALMA has measured turbulence levels too low to be consistent with MRI \citep[e.g.,][]{Flaherty2015}. However, ALMA is not sensitive to the uppermost layers of the disk and these ALMA-made turbulence measurements may only investigate the 'dead zone'; meanwhile, the MRI is active in the surface layer of the disk (see Figure~\ref{fig:MRI}).  

Given that the MRI is dependent on X-ray ionization, studies of T Tauri stars with AXIS will
better characterize the X-ray emission of young stars and allow us to connect the protostar's X-ray
luminosity to disk emission lines, which will probe the effectiveness of the MRI towards angular
momentum transport in protoplanetary disks.


\section{How stars influence their planets}
\label{sec:activity}

The structure and evolution of planetary atmospheres are profoundly affected by the high-energy irradiation and stellar wind environment produced by their host stars. In addition to the quiescent environment, magnetically driven flares and coronal mass ejections (CMEs) can impact and alter planet atmospheres via stripping and charge-exchange interactions, as observed in our own Solar System (e.g.,
Mars' atmospheric stripping and the MAVEN spacecraft's results; [REFS]). AXIS observations of stars will, directly and indirectly, probe all these physical processes, which are key factors for understanding the observed demographics of exoplanet systems and assessing the habitability of distant worlds. 

The next decade will usher in a new generation of visible and infrared instrumentation for the detection and characterization of exoplanets. This ever-growing parameter space demands rapid and reliable estimates of the space weather environment, for the purposes of identifying prime targets for atmospheric characterization and habitability studies. 
When possible, combined FUV and X-ray observations of
planet-hosting stars can be used to derive the total XUV flux that is primarily responsible for
photoevaporation, either through a piece-wise model reconstruction or via the scaling relation between the
X-ray and the EUV flux \citep{Linsky2014, King2018}.
However, due to the high levels of extinction that the ISM poses for FUV and EUV observations, soft X-rays are the only means for probing stellar activity for the vast majority of stars that are more than 100 parsecs away. Those AXIS observations can then be used to derive the EUV flux that is primarily responsible for photoevaporation, either through a piecewise model reconstruction, or via scaling relation between the X-ray and the EUV flux \citep{Linsky2014, King2018}. When AXIS observes planet-hosting stars for which FUV measurements are available via legacy surveys, the full spectrum of the stellar EUV emission can be modeled to much higher accuracy. 

\begin{figure}[H]
\centering
\includegraphics[width=12 cm]{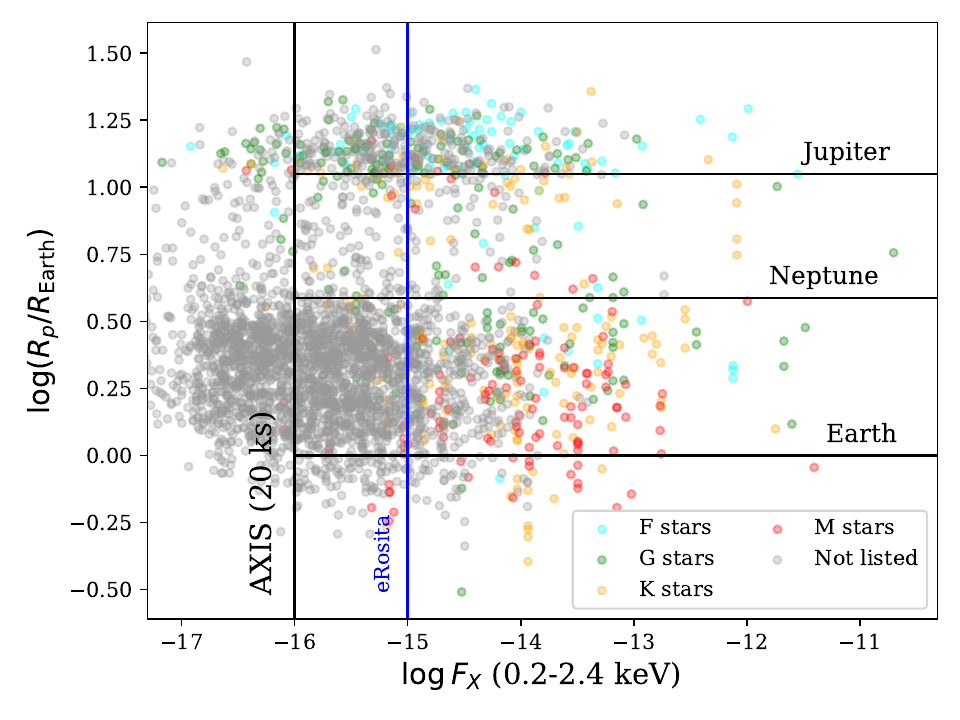}
\caption[{Predicted X-ray demographics of planet-hosting stars}]{As of this writing, there are 5438 confirmed planets cataloged in the NASA Exoplanet Archive. Of all planets in the catalog, 75\% were discovered via the transit method, and there are $\sim 6100$ planet candidates from TESS yet to be confirmed. Currently, just under 300 systems have been characterized by X-ray observations \citep{Foster2022}. AXIS will be able to detect young M3 dwarfs out to distances of 2~kpc for a $3\sigma$ detection in 20~ks (i.e., a flux of $10^{-16}$ erg/s/cm$^2$). This limiting flux will enable AXIS to measure the X-ray luminosity, and hence the high energy environment, of over 1800 transiting planet-hosting systems across a large range of stellar types.}
\label{fig:exopops}
\end{figure}

For all these reasons, AXIS will become the prime observatory for studying the high-energy properties of stars and their magnetic activity.
AXIS will be able to characterize the space weather environment of over 1800 known planetary systems.
Figure~\ref{fig:exopops} demonstrates the leap in discovery space that AXIS will be able to achieve when it comes to studying the high-energy properties of exoplanet-hosting systems. Planet radius is plotted against the X-ray fluxes expected for 2639 stellar systems with transiting exoplanets, for which enough information is present in the NASA Exoplanet Archive \citep{Akeson2013} to estimate the X-ray luminosity from known age-rotation-activity relations or from a select sample of X-ray measurements from the literature. The vertical black line marks the AXIS flux sensitivity limit, which is an order of magnitude better than the dimmest planet host star detected in a recent eRosita study \citep{Foster2022}. 

\subsection{Stellar activity has the power to sculpt planetary systems}

One of the most remarkable results in exoplanet science today has been the discovery of direct absorption of escaping gas from hot, close-in, short-period Jupiters and sub-Neptunes \citep{VidalMadjar2003, VidalMadjar2004, Lecavelier2010, Ehrenreich2015, Linsky2010, BenJaffel2013, Fossati2010, Spake2018}. The observed mass loss could be attributed to hydrodynamic escape caused by X-ray and extreme ultraviolet (together, XUV) radiation from the host star \citep[e.g.,][]{Lammer2003, OwenJackson2012}. XUV photons are absorbed at higher planetary altitudes, in lower-density parts of the atmosphere, where they ionize atoms and molecules. The resulting free electrons heat the gas collisionally; these processes inflate the gas, potentially leading to the formation of hydrodynamic outflows \citep{VidalMadjar2003, Yelle2004, Tian2005, Koskinen2014}. Ongoing atmospheric escape has been confirmed observationally in a handful of nearby exoplanets through the detection of escaping hydrogen via transit Lyman alpha spectroscopy and, more recently, through the detection of metastable helium in the outer atmosphere \citep[e.g.,][]{Oklopvic2018, Spake2018, Lampon2020}. 

For close-in planets (within a fraction of an AU), prolonged exposure to intense and/or flaring stellar radiation can potentially lead to the removal of a substantial fraction of the atmosphere's initial light-element (H/He) gas envelope \citep[e.g.,][]{OwenWu2013, OwenLai2018, Owen2019, Wu2019}. Assuming that the incident stellar XUV flux is primarily converted into expansion work, the instantaneous mass-loss rate  from an irradiated atmosphere can be expected to scale directly with the incident XUV flux and inversely on the average planetary density \citep{Watson1981, Erkaev2007}. In practice, numerical work has shown that the applicability of this ``energy-limited'' approximation is fairly limited and that the efficiency of stellar-driven atmospheric escape depends (strongly) on the planet's gravity, as well as the intensity of photoionizing radiation \citep{Lammer2003, Yelle2004, Tian2005, MurrayClay2009, OwenJackson2012, Erkaev2013, Salz2016, Caldiroli2022}. 

These studies indicate that hydrodynamic escape could play a significant role in shaping the observed properties of the known hot population of super-Earths and sub-Neptunes. XUV-driven atmospheric escape is proposed to explain the ``radius valley", an observed dearth of planets with radii of 1.5 to $2 R_\oplus$ \citep{Lopez2012, Fulton2017, OwenWu2017}, and potentially the ``Neptune desert’’, the observed lack of short-period Neptunes \citep{Mazeh2016, OwenLai2018}.  An understanding of photoevaporative losses is needed to decipher the full picture of planet formation and evolution, starting with our understanding of the intrinsic XUV properties of planet-hosting stars. 

\begin{figure}
\centering
\includegraphics[width=8 cm]{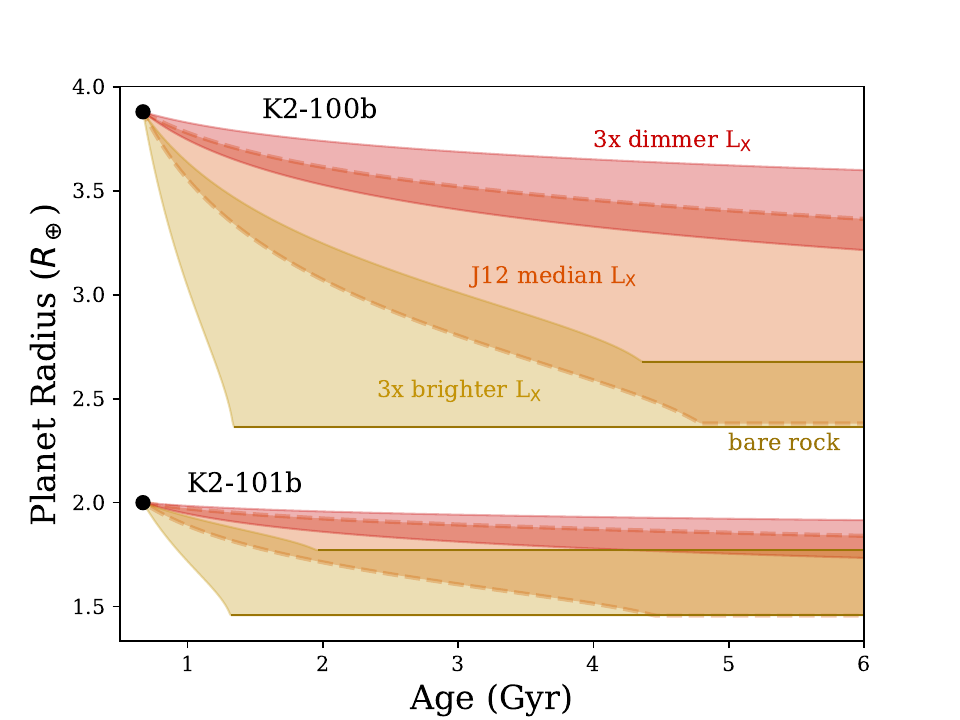}
\includegraphics[width=8 cm]{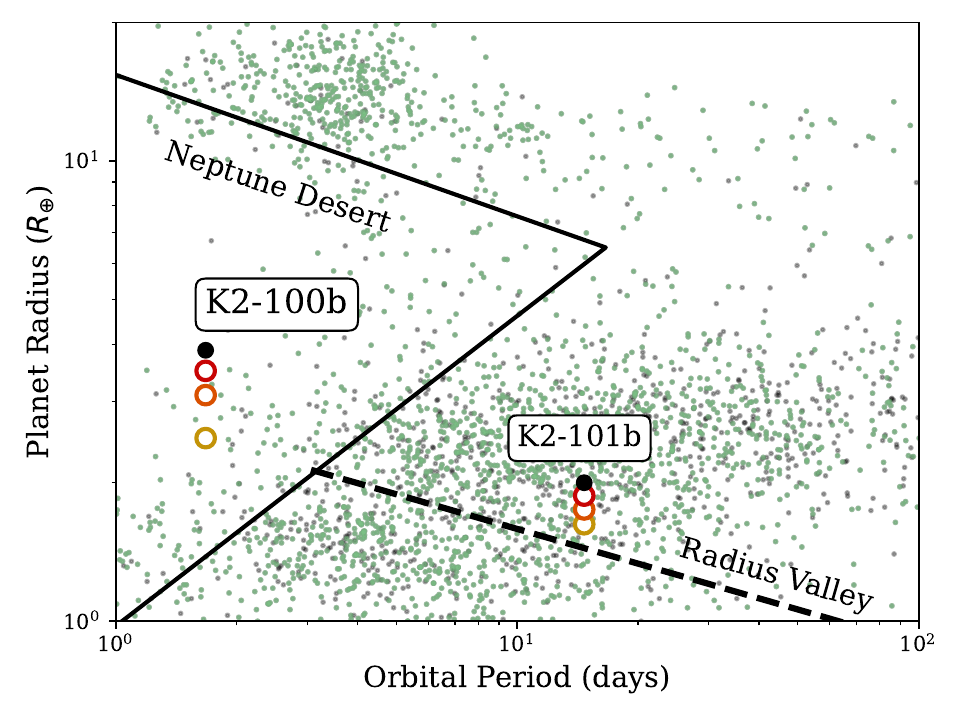}
\caption[Predicted radius evolution of exoplanets subject to photoevaporative mass loss]{\textit{(Left)} The predicted radius evolution under photoevaporative escape models for two planets in the young Praesepe star cluster \citep{King2022}. Planet K2-100b is near the edge of the Neptune desert, and planet K2-101b is a sub-Neptune near the large side of the radius-period valley. The circle markers indicate the current age and radius of these planets. The contours represent the uncertainty that arises from the uncertainty in the mass measurement of these planets. Three possible trajectories for the radius evolution of these planets is predicted, depending on whether the current-day X-ray luminosity of the host star matches the median $L_X/L_{\rm bol}$ value from the \citet{Jackson2012} relations or deviates from this value within observed dispersion of stellar X-ray luminosities. \textit{(Right)} The observed radius and orbital period of transiting planetary systems listed in the NASA Exoplanet Archive. The green dots represent planets orbiting stars that are accessible for X-ray characterization by AXIS. The solid black circles indicate the current position of the two planets on the left, and their possible positions on this diagram at age 6~Gyr are plotted with open circles.}
\label{fig:massloss}
\end{figure}

The key to reconstructing the XUV exposure histories of exoplanets is X-ray luminosity measurements of planet host stars, and only the sensitivity and imaging resolution of AXIS can enable accurate X-ray luminosity measurements across a large range of ages, for all star types down to M3V (the peak of the stellar initial mass function). From an instantaneous measurement of a star’s X-ray luminosity, its planet’s present-day XUV exposure can be calculated \citep[e.g.,][]{SanzForcada2011, Chadney2015, King2018}. Then, with an understanding of the star’s likely X-ray evolution over time, past and future XUV exposure can be probabilistically determined \citep[e.g.,][]{King2021}. Combined with theoretical models of atmospheric escape, this XUV history can be used to predict (or trace) the future (or past) evolution of individual planets’ atmospheres \citep[e.g.,][]{King2022},Figure~\ref{fig:massloss}.

Unfortunately, this process is complicated by vast uncertainties in our basic understanding of how the X-ray luminosity of FGKM-type stars evolves with age. Young stars typically exhibit $L_X/L_{\rm bol} \approx 10^{-3}$, dubbed the ``saturated'' regime, until they reach about 100~Myr to 1~Gyr in age (depending on spectral type), after which the average $L_X/L_{\rm bol}$ decays as a power law with time \citep[e.g.,][]{Jackson2012, Booth2017, Johnstone2021}. However, scattered about these average trends is a large dispersion of an order of magnitude or more in the observed X-ray luminosities, which represent a single snapshot in time. Modeling the diversity of lifetime-integrated X-ray exposures of exoplanet atmospheres requires an understanding of the degree to which this dispersion can be attributed to intrinsic physical differences among stars or is instead representative of average temporal variability (e.g., flares) exhibited randomly by all stars. 

With the imaging resolution and collecting area of AXIS, we will be able to monitor star clusters spanning Myr to Gyr ages to capture X-ray luminosities (means and dispersions) and high-energy events (flares, coronal mass ejections, accretion shocks), enabling detailed characterization of their stellar activity evolution with mass and age. AXIS has the sensitivity to reach a very wide range of stellar masses (i.e., down to the peak of the stellar IMF at spectral type around M3) at their faintest (i.e., quiescent emission at Gyr ages) and the grasp to provide robust statistical characterization of X-ray luminosities across mass and age. Monitoring these systems across the nominal mission lifetime of AXIS will sample a representative range of flare energies from $10^{33}$ to $10^{35}$ erg. These flare luminosities and occurrence rates will allow for a detailed study of host stars' impact on the formation, evolution, and fate of exoplanetary systems. 

\begin{figure}
\centering
\includegraphics[width=14 cm]{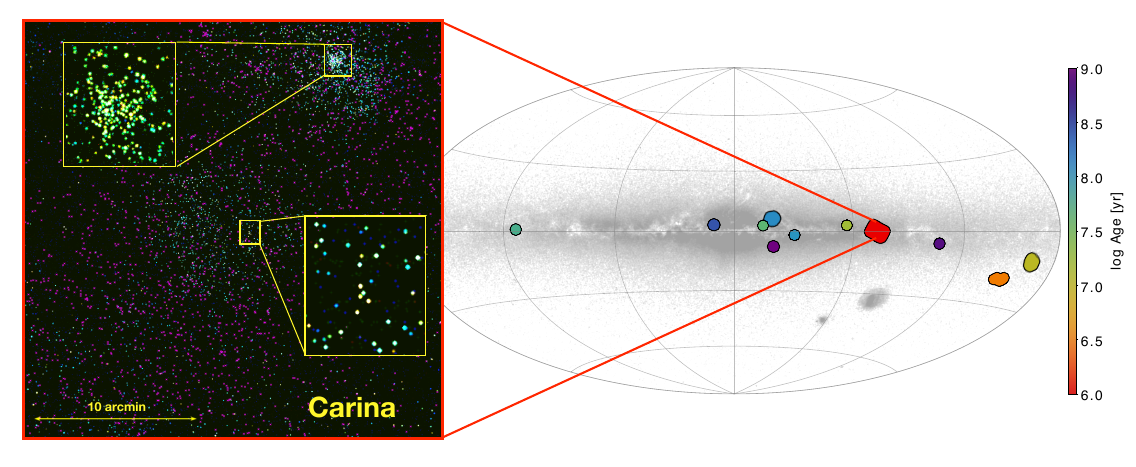}
\caption[Illustration of star clusters that AXIS can characterize]{AXIS surveys of star clusters will provide a comprehensive survey of stellar X-ray properties and the evolution of activity with age down to the peak mass of the IMF. \textit{(Left)} Simulated 50 ks AXIS observation in part of the Carina star-forming complex overlaid by the DSS optical image. Whereas Chandra did not detect the lowest-mass stars, or measure reliable hardness ratios of most stars, after 1.2~Ms and 22~pointings, AXIS will easily detect and obtain hardness ratios down to M3 dwarfs at the peak of the IMF with just a single pointing of 150~ks. \textit{(Right)} AXIS will survey clusters with Myr to Gyr ages to statistically determine the evolution of stellar activity with age, as a function of mass. Such a survey has broad implications for the evolution and habitability of exoplanetary systems.}
\label{fig:clustersurvey}
\end{figure}

The sensitivity of AXIS is such that we can track the relationships of X-ray luminosity with age for even the dimmest stars near the peak of the stellar IMF ($\sim$M3 type stars) in old open clusters up to 2.5~kpc away. Detecting quiescent emission for stars down to M3 will require exposure times $\sim 100-500$~ks (depending on distance and age), and AXIS will repeatedly observe these fields throughout its lifetime with cadences designed to precisely measure the stars' X-ray variability to eliminate the vast stochastic uncertainty in stellar activity relations. This is of particular importance for the faint M-dwarf stars, which are the most common stars in the Universe and the vast majority of terrestrial planet-hosting stars known. Figure~\ref{fig:clustersurvey} shows an all-sky image of stars detected by Gaia \citep{GaiaMission, GaiaDR2}, with known stellar clusters and associations color-coded by age (right). A simulated image of the Carina star-forming complex (left) demonstrates the superior grasp of AXIS; with a stable $1-2$~arcsec PSF across a 20~arcmin field-of-view, many thousands of young stars will be resolved.

\subsection{Stellar activity affects atmospheric retention and habitability}

A planet’s affinity for developing life as we know it requires a life-sustaining atmosphere, the longevity of which depends sensitively on the energetic radiation from its host star. Solar-type (FGKM) stars steadily produce XUV light via rotationally driven dynamos (which are mass-dependent due to the convective zone depth and stellar rotation rate both depending on stellar mass) that heat stellar chromospheres and coronae to $10^5-10^6$~K temperatures, gradually irradiating and eroding planet atmospheres over stellar lifetimes. More dramatically, powerful impulsive events (X-ray flares) and the coronal mass ejections (CMEs) associated with stellar magnetic activity can rapidly strip planetary atmospheres \citep{Vida2017, Roetten2017}, either rendering a planet barren and uninhabitable or making room for a more hospitable secondary atmosphere to accumulate \citep{OwenMohanty2016}. 
The liquid water planetary habitable zones for the most common
M dwarf stars are within $\sim$0.5 au of the host star, which is also likely to be highly X-ray active.
Understanding the evolution of stellar X-ray activity, including stellar winds, stellar flaring rates, and CME properties, as a function of age and mass is thereby required to develop a complete picture of what planetary atmospheres experience over the course of a stellar lifetime.

AXIS will contribute significantly to our understanding of habitability in our own Solar System and beyond. X-ray observations of Solar System planets yield key insights for the interaction of the Solar wind with planetary atmospheres (Section~\ref{sec:SolarSystem}). After all, it is hypothesized that the Solar wind is responsible for stripping Mars of its atmosphere, leading to the ultimate demise of its vast reservoirs of liquid surface water \citep{Lammer2003}. With the imaging resolution of AXIS, we can also seek the first detection of astrospheres (Section~\ref{sec:Astrospheres}) –- the exo-planetary system equivalent of the heliosphere -- for which spectral signatures of charge exchange will provide direct measurements of the stellar wind flux for planetary systems beyond our own. 
Astrospheres, like our heliosphere, are important because they are the environment that a star
carves out of the ISM in which planets grow and evolve.
The slew speed and timing resolution of AXIS will also be key for evaluating X-ray flares and searching for direct signatures of CMEs through coronal dimming (Section~\ref{sec:CMEs}). And finally, AXIS provides a step up in sensitivity and spectral resolution that will enable full XUV characterization of key exoplanet systems: M dwarfs hosting rocky planets in the habitable zone (Section~\ref{sec:XUV}).

\subsubsection{X-ray studies of the Nearby Solar System Illuminated by its Host Star's XUV Radiation and
Stellar Wind}
\label{sec:SolarSystem}

AXIS is uniquely poised for solar system science in a manner not possible with prior X-ray observatories. Numerous solar system objects will be observable by AXIS during the mission’s lifetime. The instrument’s large effective area will allow for investigations of temporal variations in X-ray emissions from our solar system’s planets and local planetesimals, such as asteroids and comets within ~1-2 AU of Earth. This will be the first time a thorough time-domain survey of solar system objects can be performed in X-rays. Furthermore, the fantastic spatial resolution will provide high-quality images that may be used to generate chemical composition maps of neighboring exo-systems. Utilizing both the spatial and time resolution of AXIS will allow us to probe both the transportation of elements and chemical reactions in a system due to high-energy activity, such as solar flare impacts. 

Below is a list, in no priority order, giving a number of new Solar System science programs engendered by AXIS. 
\begin{itemize}[leftmargin=*,labelsep=5.8mm]
    \item Obtain high-spatial resolution elemental composition maps of the Lunar surface.
    \item Observe high-energy particle transfer in Jupiter’s upper atmosphere, and compare the results with models built on the latest Juno results. 
    \item Follow-up on X-ray detections from the Pluto and Uranus systems to determine the origin of the signals and probe the Plutonian exosphere. 
    \item Investigate variations in comet emissions during perihelion approach, including rapid outflow events upon crossing the snow line. Determine if scattering emission is a significant X-ray emission mechanism for such systems. 
    \item Search for X-ray emission from new sources, like the Saturnian aurora, Neptune, and the asteroids.
    \item Search for and study astrospheres to understand the stellar wind fluxes,compositions, and local ISM density around other star-planet systems.
\end{itemize}
For each of these programs, in turn, we will discuss the main science drivers, expectations with AXIS, and possible limitations that AXIS may encounter. \\

\noindent \textit{Lunar Elemental Composition Map} \\

Lunar X-rays were first remotely mapped on the Apollo 15 mission \citep{Adler1972}, and are dominated by fluorescence line emissions of O, Mg, Al, and Si from the sunlit side of the Moon \citep{Schmitt1991, Wargelin2004}. Subsequent mapping of the elemental composition for the nearside Lunar surface was performed using well-characterized fluorescence line emissions, with the current highest surface resolution being 20 km from SELENE \citep{Yokota2009}. AXIS will have 10 times the spatial resolution, making it capable of obtaining a 2~km surface resolution on the abundances of O, Mg, Al, Si, and Fe on the moon using only a 100 ksec exposure. Such a resolution will allow accurate extrapolation between the remote map of AXIS to the high-resolution maps of 1-10 km regions performed from lunar orbiters/landers (e.g Lunar Reconnaissance Orbiter \citep{Chapman1994, Chin2007}). Deeper exposures will also allow AXIS to detect more exotic elements, such Au and S, that are not observable with current instruments at sufficient signal-to-noise to be properly fitted with spectral models. \\

\noindent \textit{Propagation of Martian Energetic Neutral Atoms} \\

The diffuse Martian exosphere has been known to emit brightly in X-rays since the groundbreaking study of \citet{Dennerl2002}. Martian X-ray emissions are dominated by the charge-exchange interaction between solar wind ions and neutral particles in the atmosphere, generating X-rays and energetic neutral atoms (ENAs) \citep{Krasnopolsky2002, Futaana2006, Lewkow2014}. Although the X-ray emission directly generated from this process has been detected with XMM and Chandra, thermal X-ray emission from the ENAs has yet to be remotely observed. The order-of-magnitude increase in effective area over current missions will permit AXIS to study the time-domain evolution of ENAs exiting the upper atmosphere after solar-wind ion collisions with Martian neutrals. Time domain X-ray emission studies, when combined with solar wind monitor data, will constrain cross-section models for the charge-exchange interactions, allow for the study of secondary hot atoms cross-sections and distributions, and probe the transfer of thermal energy from the exosphere to the Martian surface. \\

\noindent \textit{Variability of Jovian Magnetosphere and Exosphere} \\

X-ray emission from Jupiter’s disk, poles, and flux torus has been known since the pioneering ROSAT observations of \citet{Waite1995}. The disk emission is mainly sourced by the scattering of solar x-ray photons and carries their variability \citep{Branduardi2007, Branduardi2008}, while the polar and flux tube components are coupled to the particles and fields and "weather” of the Jovian magnetosphere and exosphere \citep{Crary1997, Connerney1998, Clarke2002}. AXIS will detect Jovian X-ray emissions with a sensitivity that is  $\sim 10\times$ deeper than Chandra, allowing detailed mapping of the source components. In particular, this will enable measurements of the emitted Io neutrals through the flux torus, testing models of the $E \times B$ deposition onto Jupiter’s poles. These results will be correlated with the latest JUNO/Chandra optical and UV maps of the system. The improved sensitivity will allow AXIS to also probe these variations at time resolutions that are not currently possible. \\

\noindent \textit{Detection and Classification of Outer Solar System X-rays} \\

Chandra detected X-ray emission from Pluto at a brightness level that was ~10x higher than estimated from standard emission mechanism models \citep{Lisse2017}. The observed spectrum indicates strong emission of soft X-rays with elemental abundances corresponding to the surface area ratio of Pluto to Charon. This suggests that either  charge-exchange or fluorescence is responsible for the emission, but the observed flux is too high for either mechanism. Follow-up observations performed with XMM have been unable to confirm these initial findings due to unfavorable geometry and the inherent low X-ray flux from Pluto. AXIS will be able to detect and resolve spatial features from Pluto within 50~ks, with a signal-to-noise ratio that would tightly constrain any proposed emission models. Deeper exposures and follow-up observations with AXIS will also provide composition and density of the upper Plutonian atmosphere as a function of time, possibly also probing the transfer of particles between Pluto and Charon. 

X-rays have potentially been detected with low signal-to-noise ratios from Uranus \citep{Dunn2021}. Saturn’s aurora and Neptune have yet to be detected in the X-ray, but their measurement could tell us a lot about the charge exchange precipitation mechanism thought to dominate X-rays from the Jovian aurora, as well as the albedo for scattering of X-rays in dense atmospheres. \\

\noindent \textit{Evolution of X-ray Emissions from Comets} \\

As a comet approaches perihelion, it absorbs solar energy and ejects neutral material from its surface outward via sublimation and localized jet-streams \citep{Wegmann2004, AHearn2011}. The neutral ejecta forms a diffuse cometary atmosphere that generates X-rays from charge exchange interactions between highly-charged, heavy solar wind ions (~0.1\% of all solar wind ions) and neutral gas in the comet's extended coma atmosphere. Cometary emissions are demonstrated to be useful for indirectly probing the charge state and solar wind composition of minor ions (i.e., those heavier than He) via fitting of comet spectra at CCD-resolution \citep{Lisse2005, Bodewits2007, Snios2016}. 

The increased observation window provided by AXIS will allow a deeper study of solar wind composition via cometary observations at a wider variety of heliospheric latitudes and longitudes. New research on cometary emissions suggests that coherent scattering of solar X-rays by cometary dust and ice particles may also contribute significantly to the total emission intensity at energies greater than 1~keV \citep{Krasnopolsky1997, Snios2014}, though deep exposures of comets with sufficient signal-to-noise spectra are required before conclusive remarks can be made on the matter. AXIS will also permit a comprehensive investigation of the presence of scattered solar X-ray emission from comets at energies greater than 1~keV, a current topic of debate in the community that is limited by the lack of signal-to-noise available with current instruments. 

Current limitations in X-ray instruments restrict cometary observations to be exclusively during a comet’s brightest activity period during their perihelion approach, allowing for only $1-2$ visits per comet. AXIS will resolve cometary emissions at distances $2-3$ times farther away than presently possible, allowing one to study the evolution X-ray emissions from comets as they pass through the snow line and begin to rapidly sublimate water from their surface. This period of activity is also when outflows of dust/ice grains of volatiles that have accumulated on the comet’s surface during its passage through the outer Solar System are ejected and interact directly with the ever-increasing solar wind flux. Probing this short-time domain of high-energy chemistry is an exciting prospect for AXIS. \\

\noindent \textit{Asteroidal X-rays} \\

X-rays have yet to be detected from asteroids, but like the lunar surface, should emit due to scattering and fluorescence of solar X-ray photons. Deep exposure observations of nearby asteroids with AXIS may provide the first detection of X-ray emissions from asteroids. Measurement of the asteroidal versus lunar albedo for scattering will inform resource usage of these bodies, as well as help solve the problem of X-ray scattering from cometary dust coma, described above.

\subsubsection{Astrospheres: direct observations of stellar wind environments of other worlds}
\label{sec:Astrospheres}

The winds of late-type stars are tenuous and very difficult to observe directly. This is a major problem for understanding the energetic stellar wind environments of planets, and also for understanding wind-driven angular momentum loss, stellar rotational evolution and the evolution of stellar magnetic activity. AXIS will be able to address this through the detection of X-ray charge-induced emission resulting
from the interaction of stellar wind ions with inflowing ISM neutrals \citet{Wargelin2001}.

Like the heliosphere of our own Solar System, all stars carve out a bubble, or ``astrosphere'', in the ISM via their outflowing stellar winds. As they travel through the galaxy, stars are surrounded by this bubble of charged gas and magnetic fields, rounded with a soft shock at the front towards the ram direction and trailing into a long tail behind. It is within the protective astrosphere that the host star dominates the local environment and planetary systems. In general, astrospheres are most easily studied at their boundaries (astropauses) where hot stellar wind ions interact with  shocks in the cold neutral ISM via highly efficient charge exchange. The  high spatial resolution and imaging capability of AXIS makes it plausible to observe astropause X-rays from  young star winds, which are expected to generate ion fluences in excess of $10^7$~protons/cm$^2$-s-ster-MeV (scaled from the X-ray flux of the sun). Astropause charge exchange will give rise to a non-thermal low-energy (300- 600 eV) signal in addition to the thermal signature of the star. If a nearby star (within $\sim 25$~pc) is surrounded by an astrosphere of order 100 AU, the emission region will exceed 1.5~arcsec and be visible to AXIS as a slightly extended source with softer, line-dominated X-rays in the surrounding halo. An attempt to detect astrospheric charge exchange emission from Proxima Centauri using Chandra was unsuccessful due to limited sensitivity \citep{Wargelin2002}, although a useful upper limit to the wind mass loss rate was obtained. 

Detection and separation of the two flux sources will only be possible with the high resolution
and large collection area capabilities of AXIS. Such observations will place constraints on wind mass
loss rates and will probe the space weather of exoplanets.

\subsubsection{Spotting coronal mass ejections via coronal dimming}
\label{sec:CMEs}

Magnetic reconnection events in the stellar atmosphere can lead to the ejection of stellar plasmas known as coronal mass ejections (CMEs) and flare radiation emission. Fast (minute- to hour-long) stellar flares are readily observed at optical wavelengths, recently made possible by the successful launch of transiting exoplanet surveys such as Kepler/K2 \citep{Davenport2016, Ilin2019, Ilin2021} and the Transiting Exoplanet Survey Satellite (TESS) \citep{Howard2019, Feinstein2020, Gunther2020}. However, multi-wavelength monitoring campaigns utilizing X-ray observations show that X-ray flares are not always correlated with the UV and optical  \citep{Osten:2005}, and empirically establishing the relationship between flares observed in XUV and coronal mass ejections remains elusive for stars other than the Sun \citep{Aarnio:2011}. 

Nonetheless, multi-wavelength campaigns have revealed key insights into flare-affiliated mechanisms, such as coronal dimming \citep{Hudson1996, Sterling1997, Veronig2021, Loyd2022}, the Orrall-Zirker effect \citep{Orrall1976, Woodgate1992, Robinson1993}, and the Neupert effect \citep{Neupert1968, Dennis1993, Tristan2023}. The Orrall-Zirker effect is expressed as an enhancement in Ly-$\alpha$ emission during flares as the result of non-thermal proton beams, and is observable in the UV only. However, coronal dimming and the Neupert Effect can both be observed in the UV and X-ray. Unfortunately, the strong absorption of UV light by the interstellar medium makes observations of such flares nearly impossible beyond the local Milky Way, whereas AXIS will provide access to stellar flares out to several kpc.

Not every X-ray flare is associated with a CME, and determining whether a planet is subject to high-energy irradiation versus high-energy particles is essential for assessing surface habitability. Coronal dimming provides the means to determine whether an X-ray flare is also associated with a CME event: when the CME ejects coronal plasma, it leaves behind a low-density coronal hole, reducing the X-ray emission from that area. Evidence of coronal dimming can be found in light curves when there is a post-flare decrease in the overall flux of the star. Coronal dimming can also be traced by looking for emission features from  $T \sim 10^6$~K plasmas, which roughly corresponds to coronal temperatures \citep{Dissauer2018}. More recent studies have systematically searched EUV and X-ray observations with the Extreme Ultra-Violet Explorer, XMM-Newton, and Chandra for coronal dimming events. To achieve this goal, there needs to be sufficient pre- and post-flare observations to be able to say if the dimming post-flare is statistically significant. 

Of the available data searched, 21 dimmings were identified in 13 different stars \citep{Veronig2021}. This has yielded exciting insight into a growing field of looking for evidence of CMEs; however, this first-look analysis explores a small sample size. The overall occurrence rate of CMEs associated with flares has yet to be determined for other stars. Understanding this occurrence rate not only has implications for stellar astrophysics, but also for planet atmospheric retention and chemistry \citep{Segura2010, Youngblood2017, Tilley2019, Chen2021}.

Surveys of flare-active stars with AXIS will yield significant insights into coronal physics. The high spatial
and energy resolution, broad energy coverage, and efficient readout times will allow for detailed studies of
specific emission lines which trace the corona during flares and CMEs, which last on the order of minutes to
hours. We will be able to resolve lines such as highly ionized Fe, Mg, O, and Si, and in doing so we can isolate and study how specific species behave in the presence of extremely energetic outbursts on other stars. The large field-of-view will facilitate broad, statistical studies of X-ray flares, allowing us to constrain X-ray flare rates and energies and occurrence rates of CMEs, as compared to the large optical surveys that have been carried out thus far.

\begin{table}
\caption{Prime targets for high energy characterization}
\label{tab:planets}
\centering
\begin{tabular}{c|c|c}
    \toprule
    \textbf{Object} &  \textbf{0.5-2 keV flux} & \textbf{System notes} \\
    &  (erg/cm$^2$/s) & \\
    \midrule
    TRAPPIST-1 & $4 \times 10^{-14}$ &  Seven Earth-sized planets orbiting an M7.5 dwarf;  \\
        &  &  three are in the habitable zone; all seven are being observed by JWST \\
    \midrule
    LHS 1140 & $1 \times 10^{-14}$ &  M4.5 dwarf with one Earth-like planet (planet c) and one super-Earth (planet b); \\
        &  &  planet b targeted for characterization by JWST \\
    \midrule
    K2-18 & $^\dagger 1 \times 10^{-14}$ & M2.5 dwarf with a cool super-Earth (planet b); \\
        &  &  targeted for characterization by JWST \\
    \midrule
    TOI-1468 & $^\dagger 1 \times 10^{-14}$ & M3.0 dwarf with two moderately large rocky planets \\
    \bottomrule
    \multicolumn{3}{l}{\footnotesize{$^\dagger$Estimated from stellar activity relations versus age or rotation period}}
\end{tabular}
\end{table}

\subsubsection{Reconstructing the high energy environment of terrestrial exoplanets}
\label{sec:XUV}

M stars are the most abundant and long-lived stars in the Universe, as well as the smallest, making them the easiest systems in which to find potentially habitable terrestrial planets. Unfortunately, they are also very faint ($\sim 0.01 L_\odot$), making them the most difficult stars to characterize with modern X-ray observatories. The soft X-ray sensitivity of AXIS provides a factor of five improvement over the current abilities of XMM-Newton, and even more so than Chandra, which will make AXIS the premier instrument for studying the high-energy environments of exoplanetary systems.

Using only 0.75\% of the nominal AXIS guaranteed exoplanet time allocation mission, we will obtain exquisite high signal-to-noise X-ray spectra and light curves of four high-impact terrestrial planetary systems (Table~\ref{tab:planets}). Each system (Table~\ref{tab:planets}) hosts a planet that has been identified as the most ideal terrestrial planets for atmospheric characterization through either transmission or emission spectroscopy (TESS Atmospheric Characterization Working Group), and almost all are being explored with JWST Cycle 1 observations. A monitoring campaign of 250~ks per target over the course of the AXIS mission operation will be able to fully characterize the high-energy environment of these planets by investigating the properties of their host stars.

\begin{figure}[H]
\centering
\includegraphics[width=7 cm]{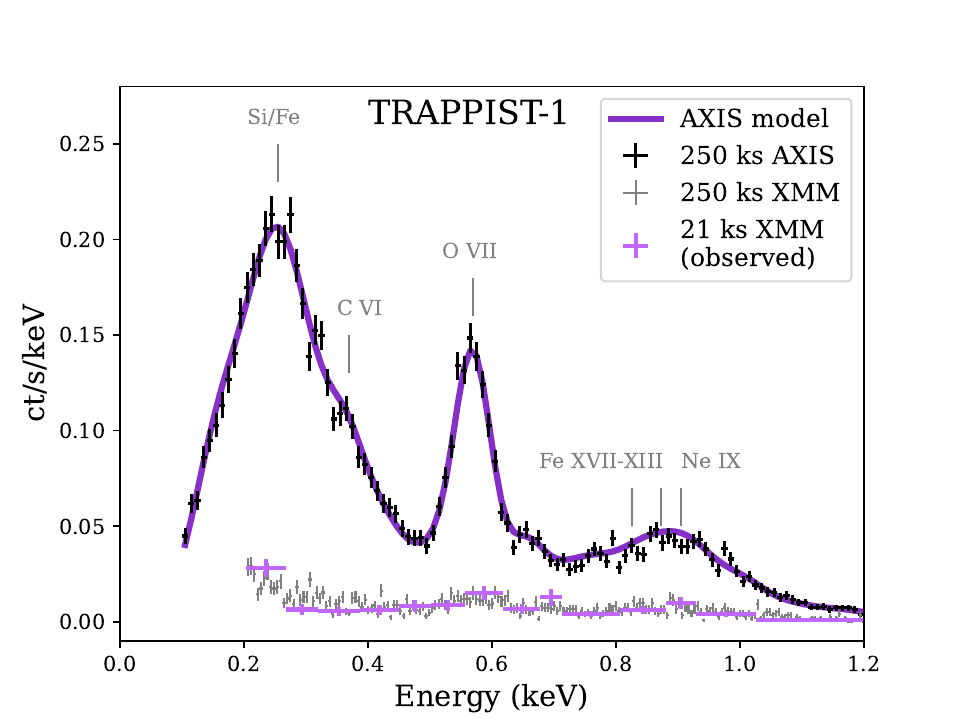}
\includegraphics[width=7.6 cm]{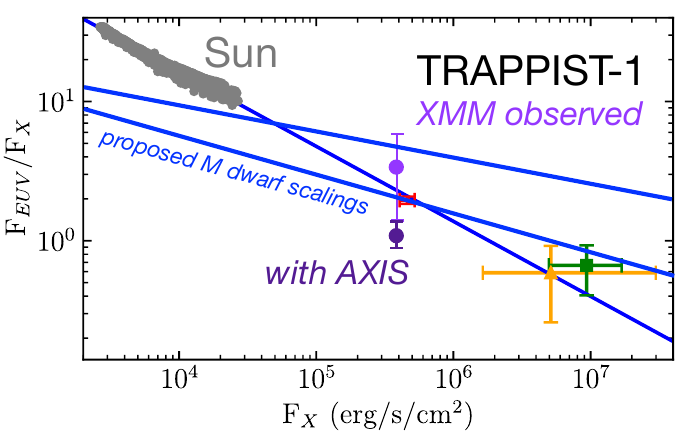}
\caption[Simulated stellar XUV modeling enabled by AXIS]{
\textit{(Left)} Simulated 250~ks spectra for the M~dwarf TRAPPIST-1 \citep{Wheatley2017} (dark purple). AXIS will provide dramatically better spectral resolution and more signal-to-noise than the same exposure time with the XMM-Newton EPIC-pn (light purple). This will provide strong constraints on the differential emission measure (DEM) reconstruction required to fully understand the stellar atmosphere and activity level.
\textit{(Right, modified from \cite{King2018})} The EUV to X-ray scaling relation inferred from Solar data is overlaid with other proposed stellar scaling relations \citep{King2018, Johnstone2021, SanzForcada2022} and the measured XUV properties of three M dwarfs (red, yellow, and gree; see \citet{Chadney2015}). The purple data points show the inferred TRAPPIST-1 EUV/X-ray luminosity ratio obtained via DEM reconstruction with a single X-ray luminosity data point (e.g., from current XMM data, light purple) versus a 250 ks observation with AXIS (dark purple). Vastly different solutions for total EUV flux are obtained when using a single X-ray luminosity as a data point versus datasets with resolved line features. The soft X-ray sensitivity and energy resolution of AXIS enhances the accuracy and lowers the EUV flux uncertainty from a factor of two to just 30\%, constraining EUV to  X-ray scaling relationships and profoundly affecting our ability to predict surface habitability of terrestrial worlds.
}
\label{fig:XUV}
\end{figure}

\noindent \textit{Temperature structure and emission measures of stellar coronae} \\

The sensitivity and spectral resolution of AXIS will provide an exquisite, high signal-to-noise spectrum from the integrated exposure of these systems. Figure~\ref{fig:XUV} shows how a 300 ks AXIS observation, with our nominal 60 eV resolution at 1 keV, can easily distinguish among different stellar coronae models. Such a spectrum will enable a full differential emission measure (DEM) solution which reconstructs the coronal plasma distribution with temperature and provides the most reliable prediction for the EUV environment of the terrestrial planet. Using the legacy quality UV dataset from the mega-MUSCLES
survey for TRAPPIST-1 \citep{Loyd2016, Wilson2021} with the AXIS data will produce 
a significant leap in accuracy of determining the EUV radiation field properties of TRAPPIST-1, reducing
model uncertainties by a factor of 3 and providing a sound XUV basis for understanding whether or not a
terrestrial-like atmosphere on TRAPPIST-1 b-h could be retained, and assessing the atmospheric chemistry and habitability of those worlds \citep{Khodachenko2007, Lammer2007, Bolmont2017, Vida2017, Wheatley2017, Johnstone2019} for such planets.  \\

\noindent \textit{Metal abundance profiles of stellar coronae} \\

Few measurements of coronal abundances are available, and many of the existing studies are at low signal-to-noise ratio. \citet{Poppenhaeger2021}  describe how EUV-driven atmospheric escape is sensitive to the coronal iron abundance, but these claims are based on very limited data. In addition to improving predictions of unobserved EUV flux (described above), measuring metal abundances in stellar coronae will resolve open questions about how magnetic fields vary across different spectral types \citep[e.g.][]{Laming2015}. Figure~\ref{fig:XUV} shows the strongest metal lines that are clearly visible in the simulated AXIS spectra.


\section{Deaths of planet-hosting stars}
\label{sec:death}

In five billion years, our Sun will enter the red giant phase, shedding its atmosphere and dramatically altering all planets in the Solar System as it makes its way to the endpoint of Solar evolution, a white dwarf. To understand the ultimate fate of Earth, we seek to characterize the plasma environment of planetary nebulae and look for evidence of planetary system remnants around other white dwarfs.

\subsection{The unique plasma environment of Planetary Nebulae}

Planetary nebulae (PNe) are proven astrophysical laboratories for a range of physical processes (shocks, ionization, photoevaporation, dust destruction, binary interactions, etc.; see \citep{Zijlstra2015, Kwitter2022}). Chandra
and XMM-Newton observations of PNe hint at a new suite of physical processes that only AXIS, with its
high-spatial resolution and increased sensitivity, is poised to answer. X-ray emission from PNe contains
multiple components, with the central soft and hard X-ray-emitting point sources \citep[][and references therein]{Montez2010, Montez2015} surrounded by nebular cavities filled with hot and diffuse X-ray emitting gas called
hot bubbles \citep[][]{Kastner2012, Ruiz2013}. The soft X-ray emission from the central star is likely due to self-shocking winds, while the hard X-ray emission stems from known and suspected spun-up binary companions \citep{Montez2015}. The diffuse hot bubble emission (filling a nebular cavity typically $5-60''$ on the sky) is intricately tied to the formation of the PN. 

During the final stages of stellar evolution, high mass loss from the slow wind of an Asymptotic Giant Branch (AGB) star is swept into the nebular shell by a new fast wind that reaches speeds of several 1000~km/s. This wind-wind interaction generates an adiabatically-shocked region with plasma temperatures that should scale with the square of the fast wind speed. However, Chandra and XMM-Newton observations show diffuse hot bubble temperatures in the 1-3 $\times 10^6$~K range instead of the $>10^7$~K predicted by their fast wind properties. AXIS will thus enable a new understanding of the physical processes required to explain the temperature discrepancy. 

\begin{figure}[H]
\centering
\includegraphics[width=14 cm]{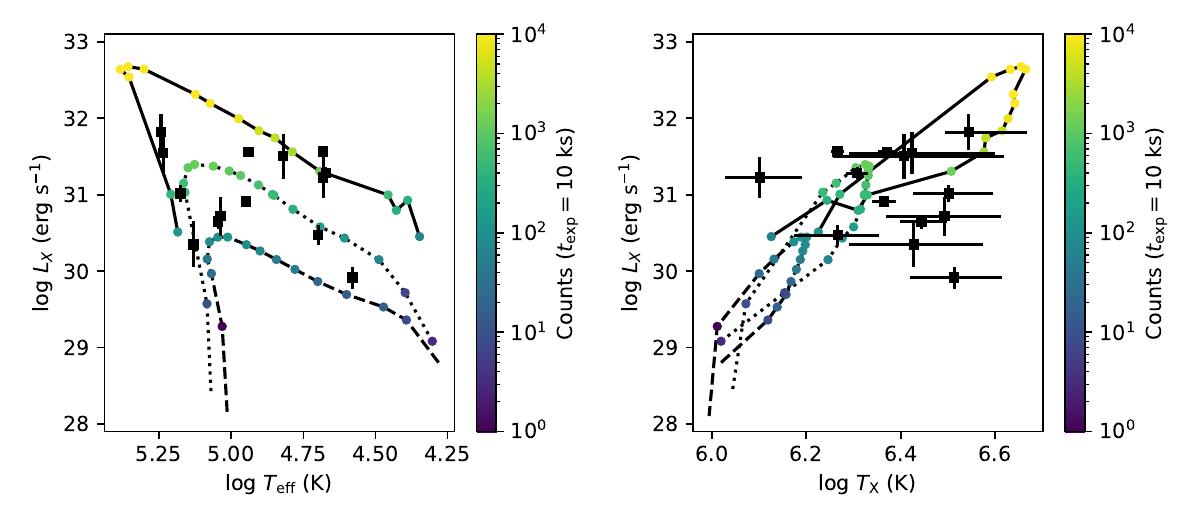}
\caption[AXIS predictions of hot bubble emission from planetary nebulae]{AXIS predictions of hot bubble emission from planetary nebulae (PNe) assuming a distance of 1.5~kpc (the ChanPlaNS distance limit). Three initial mass tracks are depicted from \citet{Steffen2008} predictions for thermal conduction in the hot bubble. The central star evolves from low $T_{\rm eff}$ to high $T_{\rm eff}$ over thousands of years. The symbols on the tracks are color-coded by the number of counts detected in 10~ks of an AXIS exposure. For comparison, the hot bubble emission from the ChanPlaNS sample is shown as black squares (Montez et al., in prep).}
\label{fig:pne}
\end{figure}

Specifically, numerical simulations suggest that clump and filament forming instabilities mix
nebular material with that of the hot bubble, producing the observed soft X-ray emission with plasma temperatures of $T_X \approx 10^6$~K \citep{Stute2006, Toala2018} and enhancing thermal conductivity \citep{Soker1994, Steffen2008,Toala2016}. Study of these processes is limited due to the faint ($L_X \approx 10^{31}$~erg/s, see Figure~\ref{fig:pne}) emission from PNe. In fact, the Chandra Planetary Nebula Survey \citep[ChanPlaNS;][]{Kastner2012, Freeman2014, Montez2015} is likely biased towards only the brightest PNe hot bubbles, which are predicted to arise from more massive progenitor stars and Wolf-Rayet-like central stars. Figure~\ref{fig:pne} demonstrates that AXIS will be able to probe hot bubble emission from lower initial mass stars for the first time as well as more distant PNe. With AXIS we will acquire high signal-to-noise images of over 100~PNe, $\sim 80$ of have been observed with weak or no signal from Chandra. Such observations will allow us to constrain the full suite of plasma properties: temperatures, chemical
abundances, and overlying absorption with unprecedented precision. Typically hot bubble emission with Chandra and XMM-Newton seldom exceeds a hundred counts; AXIS will be able to capture tens of thousands of photons from exposures as low as 10 ks. Additionally, for numerous PNe, the signal will allow us to spatially map these properties across the hot bubbles, as well as clump and filament structures. 

\subsection{The destiny of planetary systems around White Dwarfs}

It is well established that $25-50\%$ of white dwarfs exhibit metal pollution in their atmosphere, and the primary source of this pollution is thought to be the remnants of evolved exoplanetary systems \citep{Zuckerman2003, Zuckerman2010, Koester2014}. As such, these debris-accreting white dwarfs provide unique constraints on the bulk composition of exoplanetary material \citep{Zuckerman2007, Harrison2018, Hollands2018}, that serve as a key input into models of planet formation. Following the evolution and ultimate death of the host star through AGB evolution, planetary bodies (e.g., planets, moons, asteroids) entering the Roche radius are tidally disrupted, resulting in a disc of rocky debris orbiting the stellar remnant \citep{Jura2003}. As evidence of this, $2-6\%$ of polluted white dwarfs host a dusty disc \citep{Manser2020}, as inferred from an excess of infrared emission attributed to dust particle interactions in the disc \citep{Zuckerman1987, Farihi2012}. This circumstellar debris disc is believed to last for $\sim 1$~Myr, over which time the white dwarf photosphere is replenished with $Z>3$ atoms \citep{Girven2012, Cunningham2021}. Optical and UV spectroscopy enables the measurement of metal abundances, from which accretion rates, abundance ratios, and disc lifetimes can be inferred, but the scientific interpretation also requires accurate white dwarf atmospheric models. X-ray observations directly measure the accretion rate, providing an independent constraint for these models.

The recent discovery of X-rays from G29--38 confirmed for the first time the ongoing accretion of planetary debris, providing the first independent measurement of the accretion rate onto a metal-polluted white dwarf \citep{Cunningham2022}. This object was discovered via a 105 ks Chandra ACIS-S observation, which confirmed it as a source of soft X-rays with a plasma temperature of $0.5 \pm 0.2$~keV. Now that this new class of X-ray source has been discovered, it is essential to characterize the fundamental properties of these debris-accreting systems. 
 AXIS will be 150 times more sensitive to this class of objects than Chandra, providing  unprecedented ability to study the endpoints of stellar systems like our own. Firstly, a detailed follow-up of the prototypical metal-polluted white dwarf G29-38 (which will remain the stand-out target
for such studies largely owing to its proximity to Earth, 17 pc), is warranted. A 100~ks observation with AXIS will yield $\sim$1000~counts, enabling spectral features to be resolved and precision determinations of the metal abundance ratios (i.e., Mg, Si, Fe, O). 

AXIS observations will bring about a much needed connection between mid-infrared Spitzer/JWST spectroscopy (where G29--38 is a GTO target) and HST/Keck UVIS spectroscopy. Infrared spectroscopy allows a determination of the mineralogical composition of the surrounding debris disk supplying the
accreting pollutants \citep{Reach2009}. This accretion replenishes the photosphere for which UV and optical spectroscopy allows the determination of photospheric abundances \citep[e.g.,][]{Xu2014}. A spectral analysis of X-ray observations bridges the gap between these two regimes of observations. But even with the precision of JWST, there will be degeneracies in the mineralogical fitting. High signal-to-noise spectra of X-ray line emission, enabled by the soft X-ray sensitivity and energy resolution of AXIS, can break that degeneracy. 
 
 \begin{figure}[H]
\centering
\includegraphics[width=12 cm]{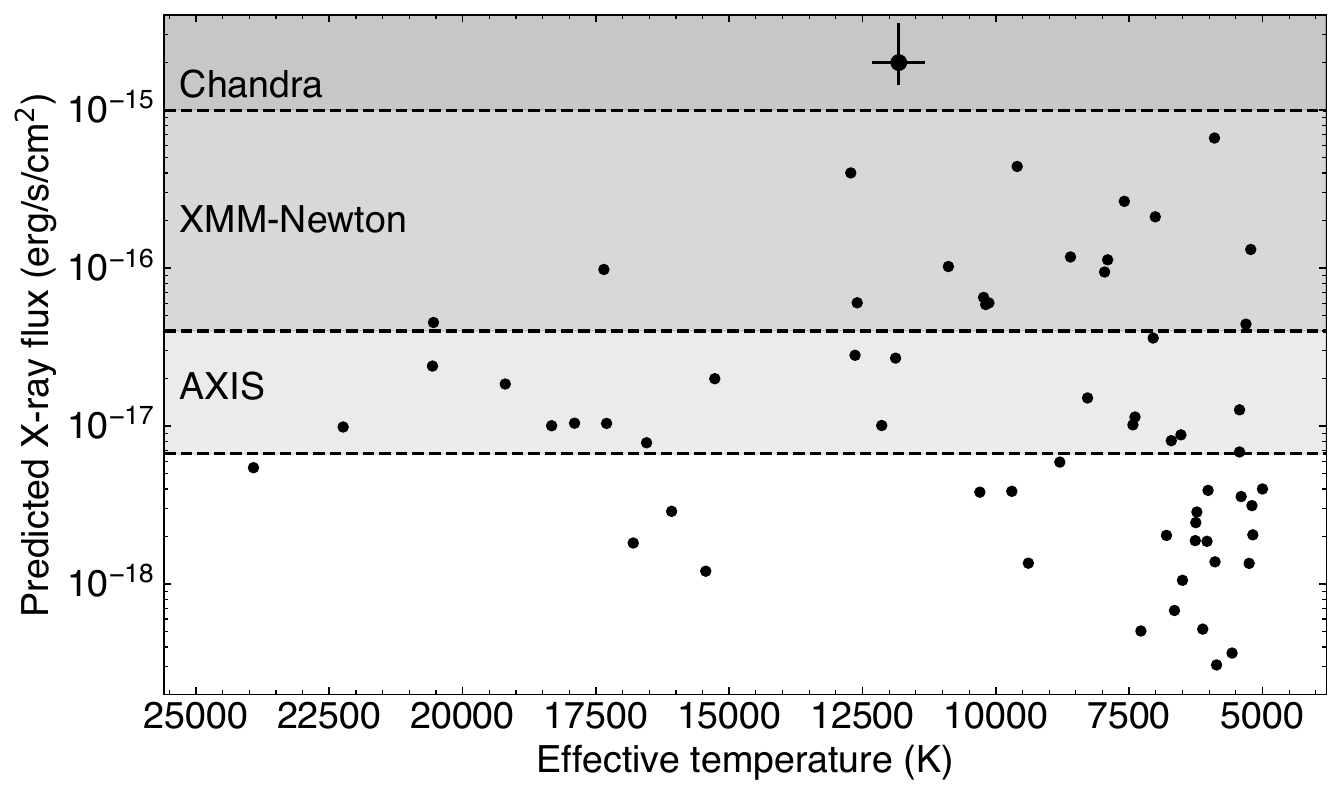}
\caption[Predicted X-ray flux of metal-polluted white dwarfs]{Predicted X-ray flux in the $0.3-7.0$~keV band for the sample of hydrogen atmosphere metal-polluted white dwarfs (DAZ) with a measured Ca abundance \citep[][Williams et al., in prep]{Farihi2016}. The X-ray flux is computed as the inferred accretion rate from optical/UV spectroscopy over the distance squared, the measured Chandra X-ray flux for G29—38 (error bars). Also indicated are estimated detection thresholds for a 115~ks exposure using Chandra, XMM-Newton, and AXIS. Thirty eight stars fall above the AXIS detection threshold, allowing accretion rates to be independently measured across the full range of effective temperatures.}
\label{fig:wds}
\end{figure}

Beyond a detailed follow-up of the closest (and currently the only known) metal-polluted white dwarf X-ray source, AXIS will enable a survey of metal-polluted white dwarfs. Figure~\ref{fig:wds} shows that a single survey of 115~ks observations should yield $30-40$ new X-ray sources. This will provide independent accretion rate measurements across a broad range of white dwarf cooling ages, allowing one to test atmospheric models throughout white dwarf evolution. In particular, there are two key uncertainties in the atmospheric models of white dwarfs: convective overshoot \citep{Freytag1996, Kupka2018, Cunningham2019} and thermohaline mixing \citep{Bauer2018, Bauer2019, Wachlin2022}. These mixing processes are of key importance across the full range of stellar evolution, and white dwarfs offer a unique window in which to test them. Both processes predict an increase in the time-averaged accretion rates from UV/optical spectroscopy because of the larger convection zone masses and longer sinking timescales. Overshoot predominantly affects DA white dwarfs cooler than 18000~K, with average accretion rates predicted to be around a factor 5 higher than models that omit overshoot \citep{Cunningham2019}, while thermohaline mixing predicts accretion rates up to 3 orders of magnitude higher for H-atmosphere white dwarfs warmer than 15000~K \citep{Bauer2019}. The increased accretion rates in these two distinct regimes can be tested with an AXIS survey of isolated white dwarfs with exposure times ranging from 10–200 ks for each white dwarf. Furthermore, differences in the mean time-integrated accretion rates of H-dominated and He-dominated atmosphere white dwarfs have been observed with spectroscopic observations (see Fig. 10 of \citep{Farihi2016}). There is no evolutionary reason that such differences should exist, and a survey of independent X-ray observations will reveal whether this is due to systematic uncertainties in the atmospheric models or if it truly is a new physical phenomenon.

\acknowledgments{This white paper represents the joint efforts and communications among the community of scientists making up the AXIS Stars \& Exoplanets Science Working Group to present the unique new contributions the AXIS mission will bring to X-ray science.}

\section*{References}
\externalbibliography{yes}
\bibliography{references, pne, cmes, Stellar, ADSlib-refs}

\end{document}